\documentclass[11pt,epsf]{article}

\begin{document}

\title{\textbf{$CP$ Violation and Family Mixing in the Effective Electroweak
Lagrangian }}
\author{D. Espriu\thanks{%
espriu@ecm.ub.es} and J. Manzano\thanks{%
manzano@ecm.ub.es} \\
Departament d'Estructura i Constituents de la Mat\`{e}ria and IFAE\\
Universitat de Barcelona, Diagonal, 647, E-08028 Barcelona}
\date{}
\maketitle

\begin{abstract}
We construct the most general effective Lagrangian of the matter sector of
the Standard Model, including mixing and $CP$ violating terms. The
Lagrangian contains the effective operators that give the leading
contribution in theories where the physics beyond the Standard Model shows
at a scale $\Lambda >>M_{W}$. We perform the diagonalization and passage to
the physical basis in full generality. We determine the contribution to the
different observables and discuss the possible new sources of $CP$
violation, the idea being to be able to gain some knowledge about new
physics beyond the Standard Model from general considerations, without
having to compute model by model. The values of the coefficients of the
effective Lagrangian in some theories, including the Standard Model, are
presented and we try to draw some general conclusions about the general
pattern exhibited by physics beyond the Standard Model in what concerns $CP$
violation. In the process we have had to deal with two theoretical problems
which are very interesting in their own: the renormalization of the CKM
matrix elements and the wave function renormalization in the on-shell scheme
when mixing is present.
\end{abstract}

\vfill
\vbox{
UB-ECM-PF 00/13\null\par
September 2000\null\par
}

\clearpage

\section{Introduction}

The origin of $CP$ violation remains, to this date, one of the unsolved
puzzles in particle physics. In the minimal Standard Model there is only one
source of $CP$ violation as is well known. Although the most general mass
matrix does, in principle, contain a large number of phases, only the left
handed diagonalization matrices survive (combined in a single
Cabibbo-Kobayashi-Maskawa, $CKM$, mixing matrix which we denote by $K$).
This matrix contains only one observable complex phase.

Whether this source of $CP$ violation is enough to explain our world is, at
present, an open question. In the near future new experimental data (mostly
involving third generation quarks) will allow us to measure with good
precision those elements of the $CKM$ matrix which are poorly known at
present. One of the commonly stated purposes of the new generation of
experiments is to check the `unitarity of the $CKM$ matrix'.

Stated this way, the purpose sounds rather meaningless. Of course if one
only retains the three known generations mixing occurs through a $3\times 3$
matrix that is, by construction, necessarily unitary. What is really meant
by the above statement is whether the observable $S$-matrix elements, which
at tree level are proportional to a $CKM$ matrix element, when measured in
charged weak decays, turn out to be in good agreement with the tree-level
unitarity relations predicted by the Standard Model. If we write, for
instance, 
\begin{equation}
\left\langle q_{j}\left| W_{\mu }^{+}\right| q_{i}\right\rangle =\mathcal{U}%
_{ij}V_{\mu }.  \label{eqa1}
\end{equation}
At tree level, it is clear that $\mathcal{U}=K$ and unitarity of the $CKM$
matrix implies 
\begin{equation}
\sum_{k}\mathcal{U}_{ik}\mathcal{U}_{jk}^{\ast }=\delta _{ij}.  \label{eqa2}
\end{equation}
However, even if there is no new physics at all beyond the Standard Model
radiative corrections contribute to the matrix elements relevant for weak
decays and spoil the unitarity of the `$CKM$ matrix' $\mathcal{U}$, in the
sense that the corresponding $S$-matrix elements are no longer constrained
to verify the above relation. Obviously, departures from unitarity due to
the electroweak radiative corrections are bound to be small. Later we shall
see at what level are violations of unitarity due to radiative corrections
to be expected.

But of course, the violations of unitarity which are really interesting are
those caused by new physics. Physics beyond the Standard Model can manifest
itself in several ways and at several scales. In this work we shall adopt
the viewpoint that new physics may appear at a scale $\Lambda $ which is
relatively large compared to the $M_{Z}$ scale. This remark includes the
scalar sector too; i.e. we assume that the Higgs particle ---if it exists at
all--- it is sufficiently heavy. If this is so, an expansion in inverse
powers of $\Lambda $ is justified and effective Lagrangian techniques\cite
{eff-lag} can be used. The scale $\Lambda $ could, for instance, be the mass
of a new heavy fermion, some compositeness scale, or simply the Higgs mass.

It is particularly interesting, at least from an instructive point of view,
to consider the case of a new heavy generation. We can proceed in two ways.
One possibility is to treat all fermions, light or heavy, on the same
footing. We would then end up with a $4\times 4$ unitary mixing matrix, the
one corresponding to the light quarks being a $3\times 3$ submatrix which,
of course need not be ---and in fact, will not be--- unitary. Stated this
way the departures from unitarity (already at tree level!) could conceivably
be sizeable. The alternative way to proceed would be, in the philosophy of
effective Lagrangians, to integrate out completely the heavy generation. One
is then left, at lowest order in the inverse mass expansion, with just the
ordinary kinetic and mass terms for light quarks, leading ---obviously--- to
an ordinary $3\times 3$ mixing matrix, which is of course unitary.
Naturally, there is no logical contradiction between the two procedures
because what really matters is the physical $S$-matrix element and this
gets, if we follow the second procedure (integrating out the heavy fields),
two type of contributions: from the lowest dimensional operators involving
only light fields and from the additional operators obtained after
integrating out the heavy fields. The result for the observable $S$-matrix
element should obviously be the same whatever procedure we follow, but using
the second method we learn that the violations of unitarity in the (three
generation) unitarity triangle are suppressed by some heavy mass (since an
additional generation decouples in the observables we are interested \cite
{hoker}) . This simple consideration illustrates the virtues of the
effective Lagrangian approach. We shall say more about this later.

The purpose of this paper is to use the philosophy behind effective
Lagrangians to try and learn some more insight on the issue of possible
sources of $CP$ violation beyond the Standard Model. We shall, in
particular, determine the most general parametrization, to the lowest
non-trivial order, of all possible family mixing and $CP$-violating effects
in the matter sector of the Standard Model. (Of course, being completely
general is impossible, so some restrictions shall apply to our
considerations. These shall be spelled out in section \ref{efflag}.)

According to our philosophy we shall, first of all, classify all possible
operators of lowest dimensionality which, respecting all the appropriate
symmetries, can be added to the ones which are present in the minimal
Standard Model. Then we shall analyze the most general kinetic and mass
terms (including, obviously, mixing). Even these terms may already be
different from those in the minimal Standard Model, the reason being that
some field redefinitions which are routinely done in the Standard Model are
not innocuous in more general models. We then proceed to diagonalize both,
the mass and kinetic terms, and determine the effects of the diagonalization
procedure, i.e. of passing to the physical basis, on the most general set of
operators of dimension four (again including the possibility of off-diagonal
couplings in family space). We then discuss the conditions for these
operators to be $CP$-odd.

Note that in the minimal Standard Model, only the left-handed
diagonalization matrices appear in physical processes (combined in the $CKM$
matrix $K$). When operators beyond the Standard Model are included
(originally written in the basis of weak eigenstates) the passage to the
physical (diagonal) basis becomes more involved. Operators involving just
left handed fields transform into more complex structures involving $K$ and
redefined effective couplings. These structures were not present before the
change of variables because, in the weak eigenstates basis, they explicitly
break $SU(2)_{L}$. For operators involving right handed fields the situation
is different. We will show that passing to the physical basis amounts only
to a redefinition of their couplings, without changing their structure. It
comes perhaps as a surprise that beyond the Standard Model the passage to
the physical basis involves in either case \textit{non-unitary} matrices.

One of the major contributions presented in this work is the detailed
treatment of the issue of wave-function and $CKM$ matrix elements
renormalization constants. There are two reasons to do so. On the one hand,
contact with physical matrix elements requires that the external legs are
properly normalized and there is a priori no reason why new physics cannot
contribute to the wave-function renormalization constants, exactly as they
do to the effective vertices. It is simply inconsistent to include one and
not the other. In fact, in the case of the $CKM$ matrix elements, their
renormalization turns out to be related to the wave-function renormalization
matrices, so it is obviously necessary to deal with this issue one way or
another even in the Standard Model. On the other hand, it must be said that
the actual on-shell prescription to incorporate the wave-function
renormalization conditions is not fully understood yet when mixing is
present. This provides for us a second motivation to treat this problem
carefully.

Another motivation to present the effective Lagrangian analysis of the
family mixing and $CP$ violation problems is that it can be applied to an
analysis of radiative corrections (for instance in the minimal Standard
Model itself) through the use of effective couplings. For a particular
process the leading contribution from radiative corrections comes as a
redefinition of the effective couplings, i.e. to some specific values for
the coefficients of the effective Lagrangian. Once determined, they can be
used for other observables without needing to compute them anew. This
procedure proved to be very efficient in recent years in the context of LEP
physics and neutral currents phenomenology \cite{bardin}.

Finally and somewhat related to the previous issue is the fact that an
effective Lagrangian provides a convenient book-keeping device to treat
deviations with respect to the Standard Model tree level predictions in a
particular process. Questions like whether is it legitimate or not to use
the unitarity of the $CKM$ matrix in a given process, given that one is
precisely looking for violations of unitarity, can be posed and answered
systematically in an effective Lagrangian framework.

The paper is organized in the following way. In the next section we extend
the effective electroweak Lagrangian in the matter sector\cite{appel} to the
case where there is mixing amongst different generations. We shall see which
restrictions $CP$-conservation imposes on the coefficients of the effective
Lagrangian. We shall then discuss in section \ref{physical} the passage to
the physical basis, which is quite interesting in the present framework, and
is in fact one of the main results of this work. The effective couplings and
some possible observable effects are discussed in section \ref{effective}.
In section \ref{renorm} we shall take into account the effects due to
renormalization, comment the expected size of the Standard Model radiative
corrections and point out some open problems. In section \ref{examples} we
shall briefly consider two examples: a heavy doublet and the Standard Model
with a heavy Higgs. Conclusions shall be summarized in section \ref
{conclusions}.

\section{Effective Lagrangian}

\label{efflag}Let us first state the assumptions behind the present
framework. We shall assume that the scale of any new physics beyond the
Standard Model is sufficiently high so that an inverse mass expansion is
granted, and we shall organize the effective Lagrangian accordingly. We
shall also assume that the Higgs field either does not exist or is massive
enough to permit an effective Lagrangian treatment by expanding in inverse
powers of its mass, $M_{H}$. In short, we assume that all as yet undetected
new particles are heavy, with a mass much larger than the energy scale at
which the effective Lagrangian is to be used. Thus it is natural to use a
non-linear realization of the $SU(2)_{L}\times U(1)_{Y}$ symmetry where the
unphysical scalar fields are collected in a unitary $2\times 2$ matrix $U$
(see e.g. \cite{eff-lag}).

An additional assumption that we may make at some point is that, whatever is
the source of $CP$-violation beyond the Standard Model, when compared to the 
$CP$ conserving part, is `small'. This statement does need qualification.
What really matters, of course, is the observable value of the $CP$
violating parameters, which are customarily calculated in the mass
eigenstate basis. On the other hand, new physics may (or may not, we do not
know for sure) appear naturally in the weak basis; i.e. with fields
transforming as irreducible representations of the gauge group. When
operators beyond the Standard Model are included they will have, in general,
a $CP$-violating and a $CP$-conserving part when written in the weak basis.
For the sake of discussion let us imagine an scenario where the origin of
fermion masses is unrelated with the physics that contributes to effective
operators beyond those already contained in the Standard Model (perhaps
because the former is associated to a very large scale). Then new physics
can be separated somehow in two parts: one part contributes to the kinetic
and Yukawa operators in the weak basis and is responsible for the known mass
structure of the matter sector; the other part contributes, again in the
weak basis, to a set of effective operators (the one described later by Eqs.(%
\ref{effec})). If we assume, for example, that the latter are totally or
almost $CP$ conserving then can have the peculiar situation that many $CP$%
-violating phases may appear in the coefficients of the effective operators
when we pass to the physical base; phases which would not be observable in
the minimal Standard Model. In short, it is conceivable that $CP$ conserving
physics triggers $CP$-violation in the physical basis. Of course the
converse is theoretically also possible, $CP$ violating phases may disappear
once things are written in the physical basis.

Let us commence our classification of the operators present in the matter
sector of the effective electroweak Lagrangian. We shall use the following
projectors 
\begin{equation}
R=\frac{1+\gamma ^{5}}{2},\qquad L=\frac{1-\gamma ^{5}}{2},\qquad \tau ^{u}=%
\frac{I+\tau ^{3}}{2},\qquad \tau ^{d}=\frac{I-\tau ^{3}}{2},  \label{eqa3}
\end{equation}
where $R$ is the right projector and $L$ the left projector in chirality
space, and $\tau ^{u}$ is the up projector and $\tau ^{d}$ the down
projector in $SU(2)$ space. The different gauge groups act on the scalar, $U$%
, and fermionic, $f_{L},f_{R}$, fields in the following way 
\begin{eqnarray}
D_{\mu }U &=&\partial _{\mu }U+ig\frac{\tau }{2}\cdot W_{\mu }U-ig^{\prime }U%
\frac{\tau ^{3}}{2}B_{\mu },  \nonumber \\
D_{\mu }^{L}f_{L} &=&\partial _{\mu }f_{L}+ig\frac{\tau }{2}\cdot W_{\mu
}f_{L}+ig^{\prime }\left( Q-\frac{\tau ^{3}}{2}\right) B_{\mu }f_{L}+ig_{s}%
\frac{\lambda }{2}\cdot G_{\mu }f_{L},  \nonumber \\
D_{\mu }^{R}f_{R} &=&\partial _{\mu }f_{R}+ig^{\prime }QB_{\mu }f_{R}+ig_{s}%
\frac{\lambda }{2}\cdot G_{\mu }f_{R}.  \label{deriv}
\end{eqnarray}

The following terms are universal. They must be present in any effective
theory whose long-distance properties are those of the Standard Model. They
correspond to the Standard Model kinetic and mass terms (we use the notation 
$\mathrm{f}$ to describe both left and right degrees of freedom
simultaneously) 
\begin{eqnarray}
\mathcal{L}_{kin}^{L} &=&i\mathrm{\bar{f}}X_{L}\gamma ^{\mu }D_{\mu }^{L}L%
\mathrm{f},  \nonumber \\
\mathcal{L}_{kin}^{R} &=&i\mathrm{\bar{f}}\left( \tau ^{u}X_{Ru}+\tau
^{d}X_{Rd}\right) \gamma ^{\mu }D_{\mu }^{R}R\mathrm{f},  \nonumber \\
\mathcal{L}_{m} &=&-\mathrm{\bar{f}}\left( U\left( \tau ^{u}\tilde{y}%
_{u}^{f}+\tau ^{d}\tilde{y}_{d}^{f}\right) R+\left( \tau ^{u}\tilde{y}%
_{u}^{f\dagger }+\tau ^{d}\tilde{y}_{d}^{f\dagger }\right) U^{\dagger
}L\right) \mathrm{f}.  \label{lm}
\end{eqnarray}
$X_{L}$, $X_{Ru}$ and $X_{Rd}$ are non-singular Hermitian matrices having
only family indices, and $\tilde{y}_{u}^{f}$ and $\tilde{y}_{d}^{f}$ are
arbitrary matrices and have only family indices too. Note that in general $%
X_{Rd}\neq X_{Ru}$, as the only restriction is gauge invariance. In the
Standard Model, these matrices can always be reabsorbed by an appropriate
redefinition of the fields (we shall see this explicitly later), so one does
not even contemplate the possibility that left and right `kinetic' terms are
differently normalized, but this is perfectly possible in an effective
theory, and the transformations required to bring these kinetic terms to the
standard form do leave some fingerprints.

In order to write the above terms in the familiar form in the Standard Model
we shall perform a series of chiral changes of variables. In general, due to
the axial anomaly, these changes will modify the $CP$ violating terms 
\begin{equation}
\mathcal{L}_{\theta }=\epsilon ^{\alpha \beta \mu \nu }\left( \theta
_{1}B_{\alpha \beta }B_{\mu \nu }+\theta _{2}W_{\alpha \beta }^{a}W_{\mu \nu
}^{a}+\theta _{3}G_{\alpha \beta }^{a}G_{\mu \nu }^{a}\right) ,  \label{eqa4}
\end{equation}
but we will not care about that here.

Notice the appearance of the unitary matrix $U$ collecting the (unphysical)
Goldstone bosons. The Higgs field ---as emphasized above--- should it exist,
has been integrated out. Since the global symmetries are non-linearly
realized the above Lagrangian is not renormalizable.

In addition to (\ref{lm}) a number of operators of dimension four should be
included in the matter sector of the effective electroweak Lagrangian. They
are, to begin with, necessary as counterterms to remove some ultraviolet
divergences that appear at the quantum level due to the non-linear nature of
(\ref{lm}). Moreover, physics beyond the Standard Model does in general
contribute to the coefficients of those operators, as it may do to $X_{L}$, $%
X_{Ru}$ $X_{Rd}$, $\tilde{y}_{u}$ and $\tilde{y}_{d}$. The dimension 4
operators can be written generically as 
\begin{eqnarray}
\mathcal{L}_{L} &=&\mathrm{\bar{f}}\gamma _{\mu }M_{L}O_{L}^{\mu }L\mathrm{f}%
+h.c.,  \nonumber \\
\mathcal{L}_{R} &=&\mathrm{\bar{f}}\gamma _{\mu }M_{R}O_{R}^{\mu }R\mathrm{f}%
+h.c.,  \label{eqa5}
\end{eqnarray}
where $M_{L}$ and $M_{R}$ are matrices having family indices only and $%
O_{L}^{\mu }$ and $O_{R}^{\mu }$ are operators of dimension one having weak
indices (u,d) only. These operators were first written by \cite{appel} in
the case where mixing between families is absent. They have been recently
considered in \cite{nomix} and \cite{top}. The extension to the
three-generation case is new.

The complete list of the dimension four operators is 
\begin{eqnarray}
\mathcal{L}_{L}^{1} &=&i\mathrm{\bar{f}}M_{L}^{1}\gamma ^{\mu }U\left(
D_{\mu }U\right) ^{\dagger }L\mathrm{f}+h.c.,  \nonumber \\
\mathcal{L}_{L}^{2} &=&i\mathrm{\bar{f}}M_{L}^{2}\gamma ^{\mu }\left( D_{\mu
}U\right) \tau ^{3}U^{\dagger }L\mathrm{f}+h.c.,  \nonumber \\
\mathcal{L}_{L}^{3} &=&i\mathrm{\bar{f}}M_{L}^{3}\gamma ^{\mu }U\tau
^{3}U^{\dagger }\left( D_{\mu }U\right) \tau ^{3}U^{\dagger }L\mathrm{f}%
+h.c.,  \nonumber \\
\mathcal{L}_{L}^{4} &=&i\mathrm{\bar{f}}M_{L}^{4}\gamma ^{\mu }U\tau
^{3}U^{\dagger }D_{\mu }^{L}L\mathrm{f}+h.c.,  \nonumber \\
\mathcal{L}_{R}^{1} &=&i\mathrm{\bar{f}}M_{R}^{1}\gamma ^{\mu }U^{\dagger
}\left( D_{\mu }U\right) R\mathrm{f}+h.c.,  \nonumber \\
\mathcal{L}_{R}^{2} &=&i\mathrm{\bar{f}}M_{R}^{2}\gamma ^{\mu }\tau
^{3}U^{\dagger }\left( D_{\mu }U\right) R\mathrm{f}+h.c.,  \nonumber \\
\mathcal{L}_{R}^{3} &=&i\mathrm{\bar{f}}M_{R}^{3}\gamma ^{\mu }\tau
^{3}U^{\dagger }\left( D_{\mu }U\right) \tau ^{3}R\mathrm{f}+h.c..
\label{effec}
\end{eqnarray}
Without any loss of generality we take the matrices in family space $%
M_{L}^{1}$, $M_{R}^{1}$, $M_{L}^{3}$ and $M_{R}^{3}$ Hermitian, while $%
M_{L}^{2}$, $M_{R}^{2}$ and $M_{L}^{4}$ are completely general. If we
require the above operators to be $CP$ conserving, the matrices $M_{L,R}^{i}$
must be real (see section \ref{effective}).

In addition to the above ones, physics beyond the Standard Model generates,
in general, an infinite tower of higher-dimensional operators with $d\geq 5$
(these operators are eventually required as counterterms too due to the
non-linear nature of the Lagrangian (\ref{lm}) ). On dimensional grounds
these operators shall be suppressed by powers of the scale $\Lambda $
characterizing new physics or by powers of $4\pi v$ ($v$ being the scale of
the breaking ---250 GeV). Therefore, if the scale of new physics is
sufficiently high the contribution of higher dimensional operators can be
neglected as compared to those of $d=4$. Of course for this to be true the
later must be non-vanishing and sizeable. Thanks to the violation of the
Appelquist-Carazzone decoupling theorem\cite{AC} in spontaneously broken
theories, this is often the case, unless the new physics is tuned so as to
be decoupling as is the case in the minimal supersymmetric Standard Model
(see e.g. \cite{DHP} for a recent discussion on this matter).

\section{Passage to the physical basis}

\label{physical}Let us first consider the operators which are already
present in the Standard Model, Eq.(\ref{lm}). The diagonalization and
passage to the physical basis are of course well known, but some
modifications are required when one considers the general case in (\ref{lm})
so it is worth going through the discussion with some detail.

We perform first the unitary change of variables 
\begin{equation}
\mathrm{f}=\left[ \tilde{V}_{L}L+\left( \tilde{V}_{Ru}\tau ^{u}+\tilde{V}%
_{Rd}\tau ^{d}\right) R\right] \mathrm{f},  \label{1}
\end{equation}
with the help of the unitary matrices $\tilde{V}_{L}$ ,$\tilde{V}_{Ru}$ and $%
\tilde{V}_{Rd}$. Hence 
\begin{equation}
\left( \tilde{y}_{u}^{f}\tau ^{u}+\tilde{y}_{d}^{f}\tau ^{d}\right)
\rightarrow \left( \tilde{V}_{L}^{\dagger }\tilde{y}_{u}^{f}\tilde{V}%
_{Ru}\tau ^{u}+\tilde{V}_{L}^{\dagger }\tilde{y}_{d}^{f}\tilde{V}_{Rd}\tau
^{d}\right) ,  \label{eqa6}
\end{equation}
and 
\begin{eqnarray}
X_{L} &\rightarrow &\tilde{V}_{L}^{\dagger }X_{L}\tilde{V}_{L}=D_{L}, 
\nonumber \\
X_{Ru} &\rightarrow &\tilde{V}_{Ru}^{\dagger }X_{Ru}\tilde{V}_{Ru}=D_{Ru}, 
\nonumber \\
X_{Rd} &\rightarrow &\tilde{V}_{Rd}^{\dagger }X_{Rd}\tilde{V}_{Rd}=D_{Rd},
\label{eqa8}
\end{eqnarray}
where $D_{L}$, $D_{Ru}$ and $D_{Rd}$ are diagonal matrices with eigenvalues
different from zero. Then, with the help of the non-unitary transformation 
\begin{equation}
\mathrm{f}\rightarrow \left[ D_{L}^{-\frac{1}{2}}L+\left( D_{Ru}^{-\frac{1}{2%
}}\tau ^{u}+D_{Rd}^{-\frac{1}{2}}\tau ^{d}\right) R\right] \mathrm{f},
\label{2}
\end{equation}
we obtain 
\begin{eqnarray}
D^{L} &\rightarrow &\left( D_{L}^{-\frac{1}{2}}\right) ^{\ast }D_{L}D_{L}^{-%
\frac{1}{2}}=I,  \nonumber \\
D_{u}^{R} &\rightarrow &\left( D_{Ru}^{-\frac{1}{2}}\right) ^{\ast
}D_{Ru}D_{Ru}^{-\frac{1}{2}}=I,  \nonumber \\
D_{d}^{R} &\rightarrow &\left( D_{Rd}^{-\frac{1}{2}}\right) ^{\ast
}D_{Rd}D_{Rd}^{-\frac{1}{2}}=I,  \label{eqa9}
\end{eqnarray}
and the matrix $\tilde{y}_{u}^{f}\tau ^{u}+\tilde{y}_{d}^{f}\tau ^{d}$
transforms into 
\begin{equation}
\left( D_{L}^{-\frac{1}{2}}\right) ^{\ast }\tilde{V}_{L}^{\dagger }\tilde{y}%
_{u}^{f}\tilde{V}_{Ru}D_{Ru}^{-\frac{1}{2}}\tau ^{u}+\left( D_{L}^{-\frac{1}{%
2}}\right) ^{\ast }\tilde{V}_{L}^{\dagger }\tilde{y}_{d}^{f}\tilde{V}%
_{Rd}D_{Rd}^{-\frac{1}{2}}\tau ^{d}\equiv y_{u}^{f}\tau ^{u}+y_{d}^{f}\tau
^{d},  \label{eqa10}
\end{equation}
where $y_{u}^{f}$ and $y_{d}^{f}$ are the Yukawa couplings. Thus, the left
and right kinetic terms can be brought to the canonical form at the sole
expense of redefining the Yukawa couplings. Since this is all there is in
the Standard Model, we see that the effect of considering the more general
coefficients for the kinetic terms is irrelevant. This will not be the case
when additional operators are considered. Fermions transform, up to this
point, in irreducible representations of the gauge group.

We now perform the unitary change of variables 
\begin{equation}
\mathrm{f}\rightarrow \left[ \left( V_{Lu}\tau ^{u}+V_{Ld}\tau ^{d}\right)
L+\left( V_{Ru}\tau ^{u}+V_{Rd}\tau ^{d}\right) R\right] \mathrm{f},
\label{3}
\end{equation}
with unitary matrices $V_{Lu}$, $V_{Ru}$, $V_{Ld}$ and $V_{Rd}$ and having
family indices only. They are chosen so that the Yukawa terms become
diagonal and definite positive (see e.g. \cite{peskin}) 
\begin{equation}
\left( V_{Lu}^{\dagger }\tau ^{u}+V_{Ld}^{\dagger }\tau ^{d}\right) \left(
y_{u}^{f}\tau ^{u}+y_{d}^{f}\tau ^{d}\right) \left( V_{Ru}\tau
^{u}+V_{Rd}\tau ^{d}\right) =d_{u}^{f}\tau ^{u}+d_{d}^{f}\tau ^{d}.
\label{eqa11}
\end{equation}
After all these transformations $\mathcal{L}_{m}$ transforms into 
\begin{equation}
\mathcal{L}_{m}=-\mathrm{\bar{f}}\left\{ \left( \tau ^{u}U+K^{\dagger }\tau
^{d}U\right) \tau ^{u}d_{u}^{f}+\left( \tau ^{d}U+K\tau ^{u}U\right) \tau
^{d}d_{d}^{f}\right\} R\mathrm{f}+h.c.,  \label{6}
\end{equation}
where $K\equiv V_{Lu}^{\dagger }V_{Ld}$ is well known
Cabibbo-Kobayashi-Maskawa matrix. Note in Eq.(\ref{6}) that when we set $U=I$
we obtain 
\begin{equation}
\mathcal{L}_{m}=-\mathrm{\bar{f}}\left( \tau ^{u}d_{u}^{f}+\tau
^{d}d_{d}^{f}\right) R\mathrm{f}+h.c.,
\end{equation}
which is a diagonal mass term. Fermions now transform in \emph{reducible}
representations of the gauge group.

The left and right kinetic terms now read 
\begin{equation}
\mathcal{L}_{kin}^{R}=i\mathrm{\bar{f}}\gamma ^{\mu }D_{\mu }^{R}R\mathrm{f},
\label{eqa12}
\end{equation}
and 
\begin{eqnarray}
\mathcal{L}_{kin}^{L} &=&i\mathrm{\bar{f}}\gamma ^{\mu }L\left\{ \partial
_{\mu }+ig^{\prime }\left( Q-\frac{\tau ^{3}}{2}\right) B_{\mu }+ig\frac{%
\tau ^{3}}{2}W_{\mu }^{3}\right.  \nonumber \\
&&\left. +ig\left( K\frac{\tau ^{-}}{2}W_{\mu }^{+}+K^{\dagger }\frac{\tau
^{+}}{2}W_{\mu }^{-}\right) +ig_{s}\frac{\lambda }{2}\cdot G_{\mu }\right\} 
\mathrm{f}.  \label{eqa13}
\end{eqnarray}
$CP$ violation is present if and only if $K\neq K^{\ast }$.

As is well known, some freedom for additional phase redefinitions is left.
If we make the replacement 
\begin{equation}
\mathrm{f}\rightarrow \left[ \left( W_{Lu}\tau ^{u}+W_{Ld}\tau ^{d}\right)
L+\left( W_{Ru}\tau ^{u}+W_{Rd}\tau ^{d}\right) R\right] \mathrm{f},
\label{eqa14}
\end{equation}
we have to change 
\begin{equation}
K=V_{Lu}^{\dagger }V_{Ld}\rightarrow W_{Lu}^{\dagger }V_{Lu}^{\dagger
}V_{Ld}W_{Ld}=W_{Lu}^{\dagger }KW_{Ld},  \label{eqa15}
\end{equation}
and 
\begin{eqnarray}
d_{u} &=&V_{Lu}^{\dagger }y_{u}^{f}V_{Ru}\rightarrow W_{Lu}^{\dagger
}V_{Lu}^{\dagger }y_{u}^{f}V_{Ru}W_{Ru}=W_{Lu}^{\dagger }d_{u}^{f}W_{Ru}, 
\nonumber \\
d_{d} &=&V_{Ld}^{\dagger }y_{d}^{f}V_{Rd}\rightarrow W_{Ld}^{\dagger
}V_{Ld}^{\dagger }y_{d}^{f}V_{Rd}W_{Rd}=W_{Ld}^{\dagger }d_{d}^{f}W_{Rd},
\label{eqa16}
\end{eqnarray}
but if we want to keep $d_{u}^{f}$ and $d_{d}^{f}$ diagonal real and
definite positive, and if we suppose that they do not have degenerate
eigenstates the only possibility for the unitary matrices $W$ is to be
diagonal with $W_{R(u,d)}=W_{L(u,d)}$. This freedom can be used, for
example, to extract five phases from $K$. After this no further
redefinitions are possible neither in the left nor in the right handed
sector.

So much for the Standard Model. Let us now move to the more general case
represented at low energies by the $d=4$ operators listed in the previous
section. We have to analyze the effect of the transformations given by Eqs.(%
\ref{1}) (\ref{2}) and (\ref{3}) (here we include in (\ref{3}) the effect of
Eq (\ref{eqa14})) on the operators (\ref{effec}). The composition of those
transformations is given by 
\begin{eqnarray}
\mathrm{f} &\rightarrow &\tilde{V}_{L}\left( D^{L}\right) ^{\frac{-1}{2}%
}\left( V_{Lu}\tau ^{u}+V_{Ld}\tau ^{d}\right) L\mathrm{f}  \nonumber \\
&&+\left( \tilde{V}_{Ru}\left( D_{u}^{R}\right) ^{\frac{-1}{2}}V_{Ru}\tau
^{u}+\tilde{V}_{Rd}\left( D_{d}^{R}\right) ^{\frac{-1}{2}}V_{Rd}\tau
^{d}\right) R\mathrm{f}  \nonumber \\
&\equiv &\left( C_{L}^{u}\tau ^{u}+C_{L}^{d}\tau ^{d}\right) L\mathrm{f}%
+\left( C_{R}^{u}\tau ^{u}+C_{R}^{d}\tau ^{d}\right) R\mathrm{f}.
\label{totalt}
\end{eqnarray}
Note that because of the presence of the matrices $D$, the matrices $C$ are
in general non-unitary. We begin with the effective operators involving left
handed fields. In this case when we perform transformation (\ref{totalt}) we
obtain 
\begin{equation}
\mathcal{L}_{L}\rightarrow \mathrm{\bar{f}}\gamma _{\mu }\mathcal{O}%
_{L}^{\mu }L\mathrm{f}+h.c.,  \label{eqa 17}
\end{equation}
with the operator $\mathcal{O}_{L}^{\mu }$ containing family and weak
indices given by 
\begin{equation}
\mathcal{O}_{L}^{\mu }=N\tau ^{u}O_{L}^{\mu }\tau ^{u}+NK\tau ^{u}O_{L}^{\mu
}\tau ^{d}  \nonumber \\
+K^{\dagger }NK\tau ^{d}O_{L}^{\mu }\tau ^{d}+K^{\dagger }N\tau
^{d}O_{L}^{\mu }\tau ^{u},  \label{7}
\end{equation}
where we have defined 
\begin{equation}
N\equiv C_{L}^{u\dagger }M_{L}C_{L}^{u}.  \label{eqa18}
\end{equation}
Thus new structures do appear involving the $CKM$ matrix $K$ and left-handed
fields. The former cannot be reduced to our starting set of operators by a
simple redefinition of the original couplings $M_{L}$.

The case of the effective operators involving right handed fields ($\mathcal{%
L}_{R}$) is, in this sense, simpler because transformation (\ref{totalt})
only redefine the matrices $M_{R}$. The operators involving right-handed
fields are 
\begin{equation}
\mathcal{L}_{R}^{p}=i\mathrm{\bar{f}}\gamma _{\mu }M_{R}^{p}O_{p}^{\mu }R%
\mathrm{f}+h.c.,  \label{eqa19}
\end{equation}
with 
\begin{equation}
O_{1}^{\mu }=U^{\dagger }\left( D_{\mu }U\right) ,\qquad O_{2}^{\mu }=\tau
^{3}U^{\dagger }\left( D_{\mu }U\right) ,\qquad O_{3}^{\mu }=\tau
^{3}U^{\dagger }\left( D_{\mu }U\right) \tau ^{3}.  \label{eqa20}
\end{equation}
Note that because of the $h.c.$ in $\mathcal{L}_{R}^{p}$ we can change $%
O_{2}^{\mu }$ by $U^{\dagger }\left( D_{\mu }U\right) \tau ^{3}$ along with $%
M_{R}^{2}$ by $M_{R}^{2\dagger }$. So under the transformation (\ref{totalt}%
) we obtain 
\[
\mathcal{L}_{R}^{p}\rightarrow i\mathrm{\bar{f}}\gamma _{\mu }\mathcal{O}%
_{pR}^{\mu }R\mathrm{f}+h.c., 
\]
with the operators $\mathcal{O}_{pR}^{\mu }$ containing family and weak
indices given by 
\begin{eqnarray}
\mathcal{O}_{pR}^{\mu } &=&C_{R}^{u\dagger }M_{R}^{p}C_{R}^{u}\tau
^{u}O_{p}^{\mu }\tau ^{u}+C_{R}^{u\dagger }M_{R}^{p}C_{R}^{d}\tau
^{u}O_{p}^{\mu }\tau ^{d}  \nonumber \\
&&+C_{R}^{u\dagger }M_{R}^{p}C_{R}^{d}\tau ^{d}O_{p}^{\mu }\tau
^{d}+C_{R}^{d\dagger }M_{R}^{p}C_{R}^{u}\tau ^{d}O_{p}^{\mu }\tau ^{u},
\label{9}
\end{eqnarray}
hence 
\begin{eqnarray}
\sum_{p=1}^{3}\mathcal{L}_{R}^{p} &\rightarrow &\sum_{p=1}^{3}\left( i%
\mathrm{\bar{f}}\gamma _{\mu }^{\mu }\mathcal{O}_{pR}R\mathrm{f}+h.c.\right)
\nonumber \\
&=&\sum_{p=1}^{3}\left( i\mathrm{\bar{f}}\gamma _{\mu }\tilde{M}%
_{R}^{p}O_{p}^{\mu }R\mathrm{f}+h.c.\right) ,  \label{eqa21}
\end{eqnarray}
with 
\begin{eqnarray}
\tilde{M}_{R}^{1} &=&C_{+}^{\dagger }M_{R}^{1}C_{+}+C_{-}^{\dagger
}M_{R}^{2}C_{+}+C_{-}^{\dagger }M_{R}^{3}C_{-},  \nonumber \\
\tilde{M}_{R}^{2} &=&C_{-}^{\dagger }M_{R}^{1}C_{+}+C_{+}^{\dagger
}M_{R}^{2}C_{+}+C_{+}^{\dagger }M_{R}^{3}C_{-},  \nonumber \\
\tilde{M}_{R}^{3} &=&C_{-}^{\dagger }M_{R}^{1}C_{-}+C_{+}^{\dagger
}M_{R}^{2}C_{-}+C_{+}^{\dagger }M_{R}^{3}C_{+},  \label{eqa22}
\end{eqnarray}
where $C_{\pm }=(C_{R}^{u}\pm C_{R}^{d})/2$. Hence, transformations (\ref
{totalt}) can be absorbed by a mere redefinition of the matrices$\
M_{R}^{1}, $ $M_{R}^{2}$ and $M_{R}^{3}$.

\section{Effective couplings and CP violation}

\label{effective}After the transformations discussed in the previous section
we are now in the physical basis and in a position to discuss the physical
relevance of the couplings in the effective Lagrangian. On dimensional
grounds the contribution of all possible dimension four operators to the
vertices can be parametrized in terms of effective couplings (see e.g. \cite
{burgess}) 
\begin{eqnarray}
\mathcal{L}_{eff} &=&-g_{s}\mathrm{\bar{f}}\gamma ^{\mu }\left(
a_{L}L+a_{R}R\right) \lambda \cdot G_{\mu }\mathrm{f}  \nonumber \\
&&-e\mathrm{\bar{f}}\gamma ^{\mu }\left( b_{L}L+b_{R}R\right) A_{\mu }%
\mathrm{f}  \nonumber \\
&&-\frac{e}{2c_{W}s_{W}}\mathrm{\bar{f}}\gamma ^{\mu }\left(
g_{L}L+g_{R}R\right) Z_{\mu }\mathrm{f}  \nonumber \\
&&-\frac{e}{s_{W}}\mathrm{\bar{f}}\gamma ^{\mu }\left( h_{L}L+h_{R}R\right) 
\frac{\tau ^{-}}{2}W_{\mu }^{+}\mathrm{f}  \nonumber \\
&&-\frac{e}{s_{W}}\mathrm{\bar{f}}\gamma ^{\mu }\left( h_{L}^{\dagger
}L+h_{R}^{\dagger }R\right) \frac{\tau ^{+}}{2}W_{\mu }^{-}\mathrm{f},
\label{vert}
\end{eqnarray}
where we define 
\begin{equation}
a_{LR}=a_{LR}^{u}\tau ^{u}+a_{LR}^{d}\tau ^{d},\qquad b_{LR}=b_{LR}^{u}\tau
^{u}+b_{LR}^{d}\tau ^{d},\qquad g_{LR}=g_{LR}^{u}\tau ^{u}+g_{LR}^{d}\tau
^{d}.  \label{eqa23}
\end{equation}
After rewriting the effective operators (\ref{effec}) in the physical basis,
their contribution to the couplings $a_{R},a_{L},b_{R},\ldots $ can be found
out by setting $U=I$.

The operators involving right-handed fields give rise to ($c_{W}=g/\sqrt{%
g^{2}+g^{\prime \,2}}$ and $s_{W}=g^{\prime }/\sqrt{g^{2}+g^{\prime \,2}}$
are the cosinus and sinus of the Weinberg angle, respectively) 
\begin{eqnarray}
\sum_{p=1}^{3}\mathcal{L}_{R}^{p} &=&-\mathrm{\bar{f}}\gamma ^{\mu }\left( 
\tilde{M}_{R}^{1}+\tilde{M}_{R}^{2}\tau ^{3}\right) \left[ \frac{e}{s_{W}}%
\left( \frac{\tau ^{-}}{2}W_{\mu }^{+}+\frac{\tau ^{+}}{2}W_{\mu
}^{-}\right) +\frac{e}{c_{W}s_{W}}\frac{\tau ^{3}}{2}Z_{\mu }\right] R%
\mathrm{f}  \nonumber \\
&&-\mathrm{\bar{f}}\gamma ^{\mu }\tilde{M}_{R}^{3}\tau ^{3}\left[ \frac{e}{%
s_{W}}\left( \frac{\tau ^{-}}{2}W_{\mu }^{+}+\frac{\tau ^{+}}{2}W_{\mu
}^{-}\right) +\frac{e}{c_{W}s_{W}}\frac{\tau ^{3}}{2}Z_{\mu }\right] \tau
^{3}R\mathrm{f}+h.c..  \label{l0fv}
\end{eqnarray}
For the operators involving left-handed fields we have instead 
\begin{eqnarray}
\mathcal{L}_{L}^{1} &=&\bar{\mathrm{\ f}}\gamma ^{\mu }\left\{ \frac{e}{%
c_{W}s_{W}}\left( N^{1}\frac{\tau ^{u}}{2}-K^{\dagger }N^{1}K\frac{\tau ^{d}%
}{2}\right) Z_{\mu }\right.  \nonumber \\
&&+\left. \frac{e}{s_{W}}\left( N^{1}K\frac{\tau ^{-}}{2}W_{\mu
}^{+}+K^{\dagger }N^{1}\frac{\tau ^{+}}{2}W_{\mu }^{-}\right) \right\} L%
\mathrm{\ f}+h.c.,  \label{l1fv} \\
\mathcal{L}_{L}^{2} &=&-\bar{\mathrm{\ f}}\gamma ^{\mu }\left\{ \frac{e}{%
c_{W}s_{W}}\left( N^{2}\frac{\tau ^{u}}{2}+K^{\dagger }N^{2}K\frac{\tau ^{d}%
}{2}\right) Z_{\mu }\right.  \nonumber \\
&&+\left. \frac{e}{s_{W}}\left( -N^{2}K\frac{\tau ^{-}}{2}W_{\mu
}^{+}+K^{\dagger }N^{2}\frac{\tau ^{+}}{2}W_{\mu }^{-}\right) \right\} L%
\mathrm{f}+h.c.,  \label{l3fv} \\
\mathcal{L}_{L}^{3} &=&-\bar{\mathrm{f}}\gamma ^{\mu }\left\{ \frac{e}{%
c_{W}s_{W}}\left( N^{3}\frac{\tau ^{u}}{2}-K^{\dagger }N^{3}K\frac{\tau ^{d}%
}{2}\right) Z_{\mu }\right.  \nonumber \\
&&+\left. \frac{e}{s_{W}}\left( -N^{3}K\frac{\tau ^{-}}{2}W_{\mu
}^{+}-K^{\dagger }N^{3}\frac{\tau ^{+}}{2}W_{\mu }^{-}\right) \right\} L%
\mathrm{f}+h.c..  \label{l4fv}
\end{eqnarray}

The contribution from $\mathcal{L}_{L}^{4}$ is a little bit different and
deserves some additional comments. Let us first see how this effective
operator looks in the physical basis and after setting $U=I$ 
\begin{eqnarray}
\mathcal{L}_{L}^{4} &=&-\bar{\mathrm{f}}\gamma ^{\mu }\left\{ \left(
N^{4}\tau ^{u}-K^{\dagger }N^{4}K\tau ^{d}\right) \left[ -i\partial _{\mu
}+eQA_{\mu }\right. \right.  \nonumber \\
&&+\left. \frac{e}{c_{W}s_{W}}\left( \frac{\tau ^{3}}{2}-Qs_{W}^{2}\right)
Z_{\mu }+g_{s}\frac{\lambda }{2}\cdot G_{\mu }\right]  \nonumber \\
&&+\left. \frac{e}{s_{W}}\left( N^{4}K\frac{\tau ^{-}}{2}W_{\mu
}^{+}-K^{\dagger }N^{4}\frac{\tau ^{+}}{2}W_{\mu }^{-}\right) \right\} L%
\mathrm{f}+h.c..  \label{l7fv}
\end{eqnarray}
One sees that $\mathcal{L}_{L}^{4}$ is the only operator potentially
contributing to the gluon and photon effective couplings. This is of course
surprising since both the photon and the gluon are associated to currents
which are exactly conserved and radiative corrections (including those from
new physics) are prohibited at zero momentum transfer. However one should
note that the effective couplings listed in (\ref{vert}) are not directly
observable yet because one must take into account the renormalization of the
external legs. In fact $\mathcal{L}_{L}^{4}$ is the only operator that can
possibly contribute to such renormalization at the order we are working.
This issue will be discussed in detail in the next section. When the
contribution from the external legs is taken into account one observes that $%
\mathcal{L}_{L}^{4}$ can be eliminated altogether from the neutral gauge
bosons couplings (and this includes the $Z$ couplings where the conserved
current argument does not apply).

Another way of seeing this (as pointed out in \cite{nomix}) is by realizing
that after use of the equations of motion $\mathcal{L}_{L}^{4}$ transforms
into a Yukawa term, so the effect of $\mathcal{L}_{L}^{4}$ can be absorbed
by a redefinition of the fermion masses and the $CKM$ matrix, if the
fermions are on-shell, as it will be the case in the present discussion.
Then it is clear that $\mathcal{L}_{L}^{4}$ may possibly contribute to the
renormalization of the $CKM$ matrix elements only (i.e. to the charged
current sector).

All this considered, from Eqs.(\ref{vert}) and (\ref{l0fv}-\ref{l7fv}), and
from the results presented in the next section concerning wave function
renormalization, we obtain for the photon and gluon couplings 
\begin{equation}
a_{L}=a_{R}=b_{L}=b_{R}=0,  \label{eqa24}
\end{equation}
both for the up and down components. For the $Z$ couplings we get 
\begin{eqnarray}
g_{L}^{u} &=&-N^{1}-N^{1\dagger }+N^{2\dagger }+N^{2}+N^{3}+N^{3\dagger }, 
\nonumber \\
g_{L}^{d} &=&K^{\dagger }\left( N^{1}+N^{1\dagger }+N^{2\dagger
}+N^{2}-N^{3}-N^{3\dagger }\right) K,  \nonumber \\
g_{R}^{u} &=&\tilde{M}_{R}^{1}+\tilde{M}_{R}^{1\dagger }+\tilde{M}_{R}^{2}+%
\tilde{M}_{R}^{2\dagger }+\tilde{M}_{R}^{3}+\tilde{M}_{R}^{3\dagger }, 
\nonumber \\
g_{R}^{d} &=&\tilde{M}_{R}^{2}+\tilde{M}_{R}^{2\dagger }-\tilde{M}_{R}^{1}-%
\tilde{M}_{R}^{1\dagger }-\tilde{M}_{R}^{3}-\tilde{M}_{R}^{3\dagger }.
\label{Zcoupl}
\end{eqnarray}
The contribution from wave-function renormalization cancels the dependence
from the vertices on the Hermitian combination $N^{4}+N^{4\dagger }$, which
is the only one that appears from the vertices themselves.

As for the effective $W$ couplings we give next the contribution coming from
the vertices only. Naturally, in order to get the full effective couplings
one must still add the contribution from wave-function renormalization and
from the renormalization of the $CKM$ matrix elements. Actually we will see
in subsection (\ref{dk}) that these contributions cancel each other at tree
level so in fact the following results include the full dependence on $N^{4}$
\begin{eqnarray}
h_{L} &=&\left( -N^{1}-N^{1\dagger }+N^{2}-N^{2\dagger }-N^{3}-N^{3\dagger
}+N^{4}-N^{4\dagger }\right) K,  \nonumber \\
h_{R} &=&\left( \tilde{M}_{R}^{1}+\tilde{M}_{R}^{1\dagger }+\tilde{M}%
_{R}^{2}-\tilde{M}_{R}^{2\dagger }-\tilde{M}_{R}^{3}-\tilde{M}_{R}^{3\dagger
}\right) .  \label{vertices}
\end{eqnarray}

The above effective couplings thus summarize all effects due to the mixing
of families in the low energy theory caused by the presence of new physics
at some large scale $\Lambda$. Let us now investigate the possible new
sources of $CP$ violation in the above effective couplings.

Generically we can write 
\begin{equation}
\mathcal{L}_{L}=\mathrm{\bar{f}}\gamma _{\mu }S^{\mu }L\mathrm{f}+h.c.,
\label{eqa25}
\end{equation}
where 
\begin{equation}
S^{\mu }\equiv N\tau ^{u}O^{\mu }\tau ^{u}+NK\tau ^{u}O^{\mu }\tau
^{d}+K^{\dagger }NK\tau ^{d}O^{\mu }\tau ^{d}+K^{\dagger }N\tau ^{d}O^{\mu
}\tau ^{u},  \label{eqa26}
\end{equation}
then under $CP$ we have 
\begin{equation}
\mathcal{L}_{L}\rightarrow \mathrm{\bar{f}}\gamma _{\mu }S^{\prime \mu }L%
\mathrm{f},  \label{eqa27}
\end{equation}
with 
\begin{equation}
S^{\prime \mu }\equiv N^{t}\tau ^{u}O^{\mu }\tau ^{u}+K^{t}N^{t}\tau
^{d}O^{\mu }\tau ^{u}+K^{t}N^{t}K^{\ast }\tau ^{d}O^{\mu }\tau
^{d}+N^{t}K^{\ast }\tau ^{u}O^{\mu }\tau ^{d},  \label{eqa28}
\end{equation}
so in order to have $CP$ invariance we require 
\begin{eqnarray}
N &=&N^{\ast },  \nonumber \\
NK &=&NK^{\ast },  \nonumber \\
K^{t}NK^{\ast } &=&K^{\dagger }NK,  \label{eqa29}
\end{eqnarray}
which can be fulfilled requiring 
\begin{equation}
N=N^{\ast },\qquad K=K^{\ast }.  \label{eqa30}
\end{equation}
Note that this last condition is sufficient but not necessary, however if we
ask for $CP$ invariance of the complete Lagrangian (as we should) the last
condition is both sufficient and necessary. Analogously, the right-handed
contribution, given by Eq.(\ref{l0fv}), is $CP$ invariant provided 
\begin{equation}
\tilde{M}_{R}^{p}=\tilde{M}_{R}^{p\ast }.  \label{eqa31}
\end{equation}

Eqs (\ref{eqa24}), (\ref{Zcoupl}) and (\ref{vertices}) thus summarize the
contribution from dimension four operators to the observables. In addition
there will be contributions from other higher dimensional operators, such as
for instance dimension five ones (magnetic moment-type operators for
example). We expect these to be small in theories such as the ones we are
considering here. The reason is that we assume a large mass gap between the
energies at which our effective Lagrangian is going to be used and the scale
of new physics. This automatically suppresses the contribution of higher
dimensional operators. However, non-decoupling effects may be left in
dimension four operators, which may depend logarithmically in the scale of
the new physics. The clearest example of this is the Standard Model itself.
Since the Higgs is there an essential ingredient in proving the
renormalizability of the theory, removing it induces new divergences which
eventually manifest themselves as logarithms of the Higgs mass. This
enhances (for a relatively heavy Higgs) the importance of the $d=4$
coefficients, albeit in the Standard Model they are small (except for the
top) nonetheless since the $\log M_{H}^{2}/M_{W}^{2}$ is preceded by a
prefactor $y^{2}/16\pi ^{2}$, where $y$ is a Yukawa coupling (see \cite
{nomix}).

Apart from the issue of wave-function and $CKM$ renormalization, to which we
shall turn next, we have finished our theoretical analysis and we can start
drawing some conclusions.

One of the first things one observes is that there are no anomalous photon
or gluon couplings, diagonal or not in flavour. This excludes the appearance
from new physics contributions to the effective couplings and observables
considered here involving the photon and the gluon. As we have seen this can
be understood on rather general grounds but it is still nice to see it
explicitly.

We also observe at once that many complex phases appear (or disappear) in
the coefficients of the effective operators after the passage to the
physical basis. Even if the matrices $M_{L,R}$ were real (and thus the
effective operators themselves preserved $CP$) phases do appear after the
diagonalization, both due to the appearance of the usual $CKM$ matrix in
those effective operators involving left-handed fields, but also because the
diagonalization matrices appear explicitly, both for left and right-handed
operators. Furthermore the effective operators couplings are redefined by
matrices which are not unitary in general. It is conceivable that this might
enhance slightly the $CP$ violation induced by the effective operators, for
instance very large custodially breaking contributions in the new physics
(provided that these evade the rather stringent bounds coming from the $\rho 
$ parameter \cite{rho}) would give rather different values to the matrices $%
X_{Ru}$ and $X_{Rd}$, yielding eigenvalues smaller than one in one of the
two. These might enhance $CP$ violation in the right-handed sector.

In the Standard Model there is a link between the existence of three
families and the presence of $CP$ violation. This disappears completely,
both in the left and right-handed sectors, once additional operators are
included. The new $CP$-violating contributions need not, in fact, be
suppressed by the product of all the mass differences, as it happens in the
Standard Model. This is obviously so if the physics responsible for the
effective operators in the weak basis is $CP$-violating, but even if it
turns out that the new physics is such that the effective operators do not
violate $CP$ in the weak basis, both the effective left and right-handed
couplings contain many independent phases as pointed out in section \ref
{efflag}. Indeed from Eqs.(\ref{totalt}-\ref{eqa18}) we see that we can have
up to $9$ independent phases in the left sector ($1$ in $K$ and the other $8$
in the $N$'s, the latter not observable in the Standard Model) and from Eqs.(%
\ref{totalt}) and (\ref{eqa22}) we see the we can have up to $18$
independent phases in the right sector which were not observable in the
Standard Model. (See \cite{right} for some work on right-handed phases and
mixing matrices.). Obviously if the matrices $M$ are allowed to be complex
more phases are available.

How can we check for the presence of all this wealth of new phases? In the
left-handed sector the analysis is usually done in terms of the unitarity
triangle. Clearly the unitarity triangle as such is gone once the additional 
$d=4$ operators are included. To see this we need only to examine Eq.(\ref
{vertices}). The total charged current vertex will be proportional to 
\begin{equation}
\mathcal{U}=K+GK_{,}  \label{eqa32}
\end{equation}
where $G$ is a combination of the $N$ matrices. Since $G$ is not
antihermitian, $\mathcal{U}$ is not unitary in a perturbative sense. This of
course is what happens when the contribution from the new physics is
considered, but it is clear that this will happen in the Standard Model too
when radiative corrections are included, since radiative corrections give
very specific, but non-zero, values for the effective couplings which also
lead to violations of unitarity.

However, these deviations of unitarity due to radiative corrections shall be
small. We expect contributions of order $g^{2}/16\pi ^{2}$ from the gauge
sector and of order $(y^{2}/16\pi ^{2})\log M_{H}^{2}/M_{W}^{2}$ from the
scalar sector to the couplings; at most of order a few times $10^{-3}$. This
is almost certainly invisible in the ongoing generation of experiments
trying to test the $CP$-violating sector of the Standard Model. Deviations
from the tree-level predictions, expressed through the coefficients of the
effective Lagrangian and their effective coupling counterpart will
measurable at present only if they are sizably larger than the radiative
corrections themselves. It is not so easy, however, to build models where
this is so. We refer the reader to section 6 for a few more comments on
this. We would also like to draw the reader's attention to \cite{london} and
references therein.

\section{Radiative corrections and renormalization}

\label{renorm} As we mentioned in the previous section, the effective
couplings presented in (\ref{vertices}) for the charged current vertices are
not the complete story because $CKM$ and wave-function renormalization gives
a non-trivial contribution there. In this section we shall consider the
contribution to the observables due to wave-function renormalization and the
renormalization of the $CKM$ matrix elements. The issue, we shall see, is
far from trivial.

When we calculate cross sections in perturbation theory we have to take into
account the residues of the external leg propagators. The meaning of these
residues is clear when we do not have mixing. In this case, if we work in
the on-shell scheme, we can attempt to absorb these residues in the wave
function renormalization constants and forget about them. However the Ward
identities force us to set up relations between the renormalization
constants that invalidate the naive on-shell scheme \cite{Hollik}. The issue
is resolved in the following way: Take whatever renormalization scheme that
respects Ward identities and use the corresponding renormalization constants
everywhere in except for the external legs contributions. For the latter we
just have to impose the mass pole and unit residue conditions. This recipe
is equivalent to use the Ward identities-complying renormalization constants
everywhere and afterwards perform a finite renormalization of the external
fields in order to assure mass pole and residue one for the propagators.
This is the commonly used prescription in the context of the popular and
convenient on-shell scheme\cite{Hollik} and, in the context of effective
theories was used in \cite{HE} and in \cite{nomix}.

Now let us now turn to the case where we have mixing. This was studied some
time ago by Aoki et al \cite{Aoki} and a on-shell scheme was proposed.
Unfortunately the issue is not settled. We have studied the problem with
some detail anew since, as already mentioned, the contribution from
wave-function renormalization is important in the present case. We have
found out that the set of conditions imposed by Aoki et al over-determine
the renormalization constants and is in fact incompatible with the analytic
structure of the theory. Moreover, even if this problem is ignored, it was
found some time ago \cite{Grassi} that the proposal conflicts with the BRST
symmetry of the theory. Therefore, now we will analyze the renormalization
issue with some detail and then we shall propose a couple of schemes which
are free of the over-determination problem. Once we have obtained those
schemes we will show how they must be used in order to avoid conflict with
Ward identities.

The renormalized fermionic propagator is given by 
\begin{eqnarray}
S\left( p\right) &=&\frac{i}{\not{p}-m-\hat{\Sigma}\left( p\right) } 
\nonumber \\
&=&i\left( \not{p}-m-\hat{\Sigma}\left( p\right) \right) ^{-1}=i\left[
\left( 1-\hat{\Sigma}\left( p\right) \left( \not{p}-m\right) ^{-1}\right)
\left( \not{p}-m\right) \right] ^{-1}  \nonumber \\
&=&i\left( \not{p}-m\right) ^{-1}\left( 1-\left( -i\hat{\Sigma}\left(
p\right) \right) i\left( \not{p}-m\right) ^{-1}\right) ^{-1}  \nonumber \\
&=&i\left( \not{p}-m\right) ^{-1}+i\left( \not{p}-m\right) ^{-1}\left( -i%
\hat{\Sigma}\left( p\right) \right) i\left( \not{p}-m\right) ^{-1}+\cdots ,
\label{eqa33}
\end{eqnarray}
where, since we have mixing, the renormalized self energy $\hat{\Sigma}%
\left( p\right) $ have family indices. Unless explicitly said otherwise, all
expressions are valid both for up and down type fermions. We will indicate
the weak indices $u$ or $d$ only when necessary. From Poincar\'{e}
invariance we can write 
\begin{equation}
\hat{\Sigma}_{ij}\left( p\right) =\not{p}\left( \hat{\Sigma}_{ij}^{\gamma
R}\left( p^{2}\right) R+\hat{\Sigma}_{ij}^{\gamma L}\left( p^{2}\right)
L\right) +\hat{\Sigma}_{ij}^{R}\left( p^{2}\right) R+\hat{\Sigma}%
_{ij}^{L}\left( p^{2}\right) L,  \label{selfdiv}
\end{equation}
where $L$ and $R$ are left and right projectors respectively, so 
\begin{equation}
S_{ij}^{-1}\left( p\right) =-i\left( \not{p}-m-\hat{\Sigma}\left( p\right)
\right) _{ij}=-i\left( \not{p}-m_{i}\right) \delta _{ij}+i\hat{\Sigma}%
_{ij}\left( p\right) .  \label{ip}
\end{equation}
The on-shell conditions given by Aoki et al are 
\begin{eqnarray}
S_{ij}^{-1}\left( p_{j}\right) u_{j}^{s}\left( p_{j}\right) &=&0,
\label{os1} \\
\bar{u}_{i}^{s}\left( p_{i}\right) S_{ij}^{-1}\left( p_{i}\right) &=&0,
\label{os2} \\
i\left( \not{p}_{i}-m_{i}\right) ^{-1}S_{ii}^{-1}\left( p_{i}\right)
u_{i}^{s}\left( p_{i}\right) &=&u_{i}^{s}\left( p_{i}\right) ,  \label{os3}
\\
\bar{u}_{i}^{s}\left( p_{i}\right) S_{ii}^{-1}\left( p_{i}\right) i\left( 
\not{p}_{i}-m_{i}\right) ^{-1} &=&\bar{u}_{i}^{s}\left( p_{i}\right) ,
\label{os4}
\end{eqnarray}
where we do not sum over repeated indices and where $p_{i}^{2}\rightarrow
m_{i}^{2}$ (on-shell). With $u_{j}^{s}$ we indicate the Dirac spinor
satisfying the on-shell condition 
\begin{equation}
\left( \not{p}_{i}-m_{i}\right) u_{i}^{s}\left( p_{i}\right) =0.
\label{eqa34}
\end{equation}
From Eqs.(\ref{ip}) and (\ref{os1}) we obtain 
\begin{equation}
\left( \left( \hat{\Sigma}_{ij}^{\gamma L}\left( m_{j}^{2}\right) m_{j}+\hat{%
\Sigma}_{ij}^{R}\left( m_{j}^{2}\right) \right) R+\left( \hat{\Sigma}%
_{ij}^{\gamma R}\left( m_{j}^{2}\right) m_{j}+\hat{\Sigma}_{ij}^{L}\left(
m_{j}^{2}\right) \right) L\right) u_{j}^{s}\left( p_{j}\right) =0,
\label{sig1}
\end{equation}
and from there 
\begin{eqnarray}
\hat{\Sigma}_{ij}^{\gamma L}\left( m_{j}^{2}\right) m_{j}+\hat{\Sigma}%
_{ij}^{R}\left( m_{j}^{2}\right) &=&0,  \nonumber \\
\hat{\Sigma}_{ij}^{\gamma R}\left( m_{j}^{2}\right) m_{j}+\hat{\Sigma}%
_{ij}^{L}\left( m_{j}^{2}\right) &=&0.  \label{cond1}
\end{eqnarray}
Analogously from Eqs. (\ref{ip}) and (\ref{os2}) we obtain 
\begin{equation}
\bar{u}_{i}^{s}\left( p_{i}\right) \left( \left( m_{i}\hat{\Sigma}%
_{ij}^{\gamma R}\left( m_{i}^{2}\right) R+m_{i}\hat{\Sigma}_{ij}^{\gamma
L}\left( m_{i}^{2}\right) L\right) +\hat{\Sigma}_{ij}^{R}\left(
m_{i}^{2}\right) R+\hat{\Sigma}_{ij}^{L}\left( m_{i}^{2}\right) L\right) =0,
\label{eqa35}
\end{equation}
and from there 
\begin{eqnarray}
m_{i}\hat{\Sigma}_{ij}^{\gamma R}\left( m_{i}^{2}\right) +\hat{\Sigma}%
_{ij}^{R}\left( m_{i}^{2}\right) &=&0,  \nonumber \\
m_{i}\hat{\Sigma}_{ij}^{\gamma L}\left( m_{i}^{2}\right) +\hat{\Sigma}%
_{ij}^{L}\left( m_{i}^{2}\right) &=&0.  \label{cond2}
\end{eqnarray}
From Eqs. (\ref{ip}) and (\ref{os3}) we obtain 
\begin{eqnarray}
&&\hat{\Sigma}_{ii}^{\gamma R}\left( m_{i}^{2}\right) +m_{i}^{2}\left( \hat{%
\Sigma}_{ii}^{\gamma R\prime }\left( m_{i}^{2}\right) +\hat{\Sigma}%
_{ii}^{\gamma L\prime }\left( m_{i}^{2}\right) \right)  \nonumber \\
&+&m_{i}\left( \hat{\Sigma}_{ii}^{R\prime }\left( m_{i}^{2}\right) +\hat{%
\Sigma}_{ii}^{L\prime }\left( m_{i}^{2}\right) \right) =0,  \nonumber \\
&&\hat{\Sigma}_{ii}^{\gamma L}\left( m_{i}^{2}\right) +m_{i}^{2}\left( \hat{%
\Sigma}_{ii}^{\gamma L\prime }\left( m_{i}^{2}\right) +\hat{\Sigma}%
_{ii}^{\gamma R\prime }\left( m_{i}^{2}\right) \right)  \nonumber \\
&+&m_{i}\left( \hat{\Sigma}_{ii}^{L\prime }\left( m_{i}^{2}\right) +\hat{%
\Sigma}_{ii}^{R\prime }\left( m_{i}^{2}\right) \right) =0,  \label{cond3}
\end{eqnarray}
and finally from Eqs. (\ref{ip}) and (\ref{os4}) we obtain again the same
equations that we have derived from the condition (\ref{os3}). So we can
write the whole set of Aoki et al renormalization conditions as 
\begin{eqnarray}
0 &=&\hat{\Sigma}_{ij}^{\gamma L}\left( m_{j}^{2}\right) m_{j}+\hat{\Sigma}%
_{ij}^{R}\left( m_{j}^{2}\right) ,  \nonumber \\
0 &=&m_{i}\hat{\Sigma}_{ij}^{\gamma R}\left( m_{i}^{2}\right) +\hat{\Sigma}%
_{ij}^{R}\left( m_{i}^{2}\right) ,  \nonumber \\
0 &=&\hat{\Sigma}_{ii}^{\gamma R}\left( m_{i}^{2}\right) +m_{i}^{2}\left( 
\hat{\Sigma}_{ii}^{\gamma R\prime }\left( m_{i}^{2}\right) +\hat{\Sigma}%
_{ii}^{\gamma L\prime }\left( m_{i}^{2}\right) \right)  \nonumber \\
&&+m_{i}\left( \hat{\Sigma}_{ii}^{R\prime }\left( m_{i}^{2}\right) +\hat{%
\Sigma}_{ii}^{L\prime }\left( m_{i}^{2}\right) \right) ,  \label{cond+}
\end{eqnarray}
as well as those obtained by the exchange $R\leftrightarrow L$.

With the help of the mass counterterm and the left and right wave-function
renormalization constants the renormalized self energy $\hat{\Sigma}_{ij}$
can be written as 
\begin{eqnarray}
\hat{\Sigma}_{ij} &=&\Sigma _{ij}-\frac{1}{2}\not{p}L\left( \delta
Z_{ij}^{L\dagger }+\delta Z_{ij}^{L}\right) -\frac{1}{2}\not{p}R\left(
\delta Z_{ij}^{R\dagger }+\delta Z_{ij}^{R}\right)  \nonumber \\
&&+\frac{1}{2}R\left( \delta Z_{ij}^{L\dagger }m_{j}+m_{i}\delta
Z_{ij}^{R}\right) +\frac{1}{2}L\left( \delta Z_{ij}^{R\dagger
}m_{j}+m_{i}\delta Z_{ij}^{L}\right) +\delta _{ij}\delta m_{i},  \label{self}
\end{eqnarray}
where $\Sigma _{ij}$ is the bare self-energy. Using Eqs. (\ref{selfdiv}) (%
\ref{cond+}) and (\ref{self}) we can obtain the following relations among
bare self energies 
\begin{equation}
\left( \Sigma _{ij}^{\gamma L}\left( m_{j}^{2}\right) -\Sigma _{ij}^{\gamma
L\dagger }\left( m_{j}^{2}\right) \right) m_{j}+\left( \Sigma
_{ij}^{R}\left( m_{j}^{2}\right) -\Sigma _{ij}^{L\dagger }\left(
m_{j}^{2}\right) \right) =0,  \label{eqa36}
\end{equation}
and a similar relation exchanging $R\leftrightarrow L$. But we know that
this relations are not satisfied because self energies are not Hermitian
due, e.g., to the branch cut generated by the loop of massless virtual
fotons. The appearance of this type of (false) relations is due to the
over-determination of conditions (\ref{os1}-\ref{os4}).

There are several ways to solve this over-determination, here we will
present the ones that we believe are more physical.

\subsection{``Incoming fermion'' scheme}

To avoid over-determination we will define the following renormalization
conditions. We will keep for $i\neq j$ the Aoki et al renormalization
condition (\ref{os1}) namely 
\begin{equation}
S_{ij}^{-1}\left( p\right) u_{j}^{s}\left( p\right) =0\quad i\neq j,\quad
p^{2}\rightarrow m_{j}^{2},  \label{inc1}
\end{equation}
which physically means that we have no mixing on shell of the incoming
fermions and in terms of self energies amounts to 
\begin{equation}
0=\hat{\Sigma}_{ij}^{\gamma L}\left( m_{j}^{2}\right) m_{j}+\hat{\Sigma}%
_{ij}^{R}\left( m_{j}^{2}\right) ,\quad i\neq j,  \label{eqa37}
\end{equation}
and a similar condition exchanging $R\leftrightarrow L$. For $i=j$ we only
impose this condition over the real part of the inverse propagator 
\begin{equation}
Re\left( iS^{-1}\right) _{ii}\left( p\right) u_{i}^{s}\left( p\right)
=0\quad p^{2}\rightarrow m_{i}^{2},  \label{inc2}
\end{equation}
the restriction to the real part is necessary because fermions need not be
stable particles (in fact they are not in general) so an appropriate
condition for the mass pole is (\ref{inc2}), which in terms of self energies
amounts to 
\begin{equation}
0=\left( \hat{\Sigma}_{ii}^{\gamma R}\left( m_{i}^{2}\right) +\hat{\Sigma}%
_{ii}^{\gamma R\dagger }\left( m_{i}^{2}\right) \right) m_{i}+\hat{\Sigma}%
_{ii}^{L}\left( m_{j}^{2}\right) +\hat{\Sigma}_{ii}^{L\dagger }\left(
m_{i}^{2}\right) ,  \label{eqa38}
\end{equation}
and a similar condition exchanging $R\leftrightarrow L$. We also add the
unit residue condition 
\begin{equation}
\left( \not{p}-m_{i}\right) ^{-1}Re\left( iS^{-1}\right) _{ii}\left(
p\right) u_{i}^{s}\left( p\right) =u_{i}^{s}\left( p\right) ,\quad
p^{2}\rightarrow m_{i}^{2},  \label{inc3}
\end{equation}
which can be shown to be equivalent to 
\begin{equation}
\bar{u}_{i}^{s}\left( p\right) Re\left( iS^{-1}\right) _{ii}\left( p\right)
\left( \not{p}-m_{i}\right) ^{-1}=\bar{u}_{i}^{s}\left( p\right) ,\quad
p^{2}\rightarrow m_{i}^{2}.  \label{inc4}
\end{equation}
The diagonal antihermitian parts of the bare self energy are finite, so it
can be shown that in order to keep the renormalized ones finite we only need
to impose 
\begin{equation}
\delta Z_{ii}^{L}-\delta Z_{ii}^{L\dagger }=\delta Z_{ii}^{R}-\delta
Z_{ii}^{R\dagger }+\mathrm{constant}.  \label{ahpre1}
\end{equation}
In the on-shell scheme without mixing $\delta Z_{ii}^{L}-\delta
Z_{ii}^{L\dagger }=\delta Z_{ii}^{R}-\delta Z_{ii}^{R\dagger }=0$ is tacitly
assumed. However due to the rephasing freedom only condition (\ref{ahpre1})
is necessary to absorb all the divergencies. Here, for simplicity reasons,
we also take 
\begin{equation}
\delta Z_{ii}^{L}-\delta Z_{ii}^{L\dagger }=\delta Z_{ii}^{R}-\delta
Z_{ii}^{R\dagger }=0.  \label{aherii1}
\end{equation}

Note that apart from taking the real part in (\ref{inc2}-\ref{inc3}-\ref
{inc4}) we have also omitted Aoki et al condition (\ref{os2}) to avoid
over-determination. We can expect that another set of consistent condition
that include condition (\ref{os2}) (for $i\neq j$, and taking the real part
in the diagonal case) can be given, and actually this is the case.

Performing the calculations in the incoming fermion scheme we obtain the
following set of wave-function renormalization constants 
\begin{eqnarray}
\delta Z_{ij}^{L\dagger }+\delta Z_{ij}^{L} &=&\frac{2}{m_{j}^{2}-m_{i}^{2}}%
\left\{ \Sigma _{ij}^{\gamma L}\left( m_{j}^{2}\right) m_{j}^{2}-\Sigma
_{ij}^{\gamma L\dagger }\left( m_{i}^{2}\right) m_{i}^{2}\right.  \nonumber
\\
&&+\Sigma _{ij}^{\gamma R}\left( m_{j}^{2}\right) m_{i}m_{j}-\Sigma
_{ij}^{\gamma R\dagger }\left( m_{i}^{2}\right) m_{j}m_{i}  \nonumber \\
&&+\Sigma _{ij}^{R}\left( m_{j}^{2}\right) m_{j}-\Sigma _{ij}^{R\dagger
}\left( m_{i}^{2}\right) m_{i}  \nonumber \\
&&+\left. \Sigma _{ij}^{L}\left( m_{j}^{2}\right) m_{i}-\Sigma
_{ij}^{L\dagger }\left( m_{i}^{2}\right) m_{j}\right\} \quad \left( i\neq
j\right) ,  \label{set1hermLij}
\end{eqnarray}
\begin{eqnarray}
\delta Z_{ij}^{L}-\delta Z_{ij}^{L\dagger } &=&\frac{2}{m_{j}^{2}-m_{i}^{2}}%
\left\{ \Sigma _{ij}^{\gamma L}\left( m_{j}^{2}\right) m_{j}^{2}+\Sigma
_{ij}^{\gamma L\dagger }\left( m_{i}^{2}\right) m_{i}^{2}\right.  \nonumber
\\
&&+\Sigma _{ij}^{\gamma R}\left( m_{j}^{2}\right) m_{j}m_{i}+\Sigma
_{ij}^{\gamma R\dagger }\left( m_{i}^{2}\right) m_{j}m_{i}  \nonumber \\
&&+\Sigma _{ij}^{L}\left( m_{j}^{2}\right) m_{i}+\Sigma _{ij}^{L\dagger
}\left( m_{i}^{2}\right) m_{j}  \nonumber \\
&&+\left. \Sigma _{ij}^{R}\left( m_{j}^{2}\right) m_{j}+\Sigma
_{ij}^{R\dagger }\left( m_{i}^{2}\right) m_{i}\right\} \quad \left( i\neq
j\right) ,  \label{set1aherLij}
\end{eqnarray}
\begin{eqnarray}
\delta Z_{ii}^{R\dagger }+\delta Z_{ii}^{R} &=&m_{i}^{2}\left( \Sigma
_{ii}^{\gamma R\prime }\left( m_{i}^{2}\right) +\Sigma _{ii}^{\gamma R\prime
\dagger }\left( m_{i}^{2}\right) +\Sigma _{ii}^{\gamma L\prime }\left(
m_{i}^{2}\right) +\Sigma _{ii}^{\gamma L\prime \dagger }\left(
m_{i}^{2}\right) \right)  \nonumber \\
&&+m_{i}\left( \Sigma _{ii}^{R\prime }\left( m_{i}^{2}\right) +\Sigma
_{ii}^{R\prime \dagger }\left( m_{i}^{2}\right) +\Sigma _{ii}^{L\prime
}\left( m_{i}^{2}\right) +\Sigma _{ii}^{L\prime \dagger }\left(
m_{i}^{2}\right) \right)  \nonumber \\
&&+\Sigma _{ii}^{\gamma R}\left( m_{i}^{2}\right) +\Sigma _{ii}^{\gamma
R\dagger }\left( m_{i}^{2}\right) ,  \label{set1hermii}
\end{eqnarray}
and, as usual, similar conditions obtained after the exchange $%
R\leftrightarrow L$.

We also have we also have 
\begin{eqnarray}
\delta m_{i} &=&-\frac{1}{4}\left\{ \left( \Sigma _{ii}^{\gamma L}\left(
m_{i}^{2}\right) +\Sigma _{ii}^{\gamma L\dagger }\left( m_{i}^{2}\right)
+\Sigma _{ii}^{\gamma R}\left( m_{i}^{2}\right) +\Sigma _{ii}^{\gamma
R\dagger }\left( m_{i}^{2}\right) \right) m_{i}\right.  \nonumber \\
&&+\left. \Sigma _{ii}^{R}\left( m_{i}^{2}\right) +\Sigma _{ii}^{R\dagger
}\left( m_{i}^{2}\right) +\Sigma _{ii}^{L}\left( m_{i}^{2}\right) +\Sigma
_{ii}^{L\dagger }\left( m_{i}^{2}\right) \right\} .  \label{set1dm}
\end{eqnarray}
Here it is worth noting that even though this scheme has less conditions
than the Aoki et al set we still obtain restrictions over bare self
energies, namely 
\begin{equation}
\Sigma _{ii}^{L}\left( m_{i}^{2}\right) +\Sigma _{ii}^{L\dagger }\left(
m_{i}^{2}\right) =\Sigma _{ii}^{R}\left( m_{i}^{2}\right) +\Sigma
_{ii}^{R\dagger }\left( m_{i}^{2}\right) ,  \label{cons}
\end{equation}
but in this case it can be seen by direct calculation to one loop that this
relation holds.

\subsection{``Outcoming fermion'' scheme}

Another possibility is to define an on-shell scheme by the following set of
conditions. We impose 
\begin{equation}
\bar{u}_{i}^{s}\left( p\right) S_{ij}^{-1}\left( p\right) =0\quad \left(
i\neq j,\quad p^{2}\rightarrow m_{i}^{2}\right) ,  \label{out1}
\end{equation}
which physically means that we have no mixing on shell of the outcoming
fermions and in terms of self energies amounts to

\begin{equation}
0=m_{i}\hat{\Sigma}_{ij}^{\gamma R}\left( m_{i}^{2}\right) +\hat{\Sigma}%
_{ij}^{R}\left( m_{i}^{2}\right) ,  \label{eqa39}
\end{equation}
plus the $R\leftrightarrow L$ condition .

For $i=j$ we again impose this condition only over $Re\left( iS^{-1}\right) $
that is 
\begin{equation}
\bar{u}_{i}^{s}\left( p\right) Re\left( iS^{-1}\right) _{ii}\left( p\right)
=0,\quad p^{2}\rightarrow m_{i}^{2},  \label{out2}
\end{equation}
which in terms of self energies amounts to

\begin{equation}
0=\left( \hat{\Sigma}_{ii}^{\gamma R}\left( m_{i}^{2}\right) +\hat{\Sigma}%
_{ii}^{\gamma R\dagger }\left( m_{i}^{2}\right) \right) m_{i}+\hat{\Sigma}%
_{ii}^{R}\left( m_{i}^{2}\right) +\hat{\Sigma}_{ii}^{R\dagger }\left(
m_{j}^{2}\right) ,  \label{eqa40}
\end{equation}
and, as customary, the exchanged $R\leftrightarrow L$ condition. The unit
residue conditions are the same as in the incoming fermion scheme.

Performing the calculations in the outcoming fermion scheme we obtain the
following set of wave-function renormalization constants 
\begin{eqnarray}
\delta Z_{ij}^{L}+\delta Z_{ij}^{L\dagger } &=&\frac{2}{m_{i}^{2}-m_{j}^{2}}%
\left\{ \Sigma _{ij}^{\gamma L}\left( m_{i}^{2}\right) m_{i}^{2}-\Sigma
_{ij}^{\gamma L\dagger }\left( m_{j}^{2}\right) m_{j}^{2}\right.  \nonumber
\\
&&+\Sigma _{ij}^{\gamma R}\left( m_{i}^{2}\right) m_{i}m_{j}-\Sigma
_{ij}^{\gamma R\dagger }\left( m_{j}^{2}\right) m_{j}m_{i}  \nonumber \\
&&+\Sigma _{ij}^{R}\left( m_{i}^{2}\right) m_{j}-\Sigma _{ij}^{R\dagger
}\left( m_{j}^{2}\right) m_{i}  \nonumber \\
&&+\left. \Sigma _{ij}^{L}\left( m_{i}^{2}\right) m_{i}-\Sigma
_{ij}^{L\dagger }\left( m_{j}^{2}\right) m_{j}\right\} \quad \left( i\neq
j\right) ,  \label{set2hermLij}
\end{eqnarray}
\begin{eqnarray}
\delta Z_{ij}^{L}-\delta Z_{ij}^{L\dagger } &=&\frac{2}{m_{j}^{2}-m_{i}^{2}}%
\left\{ m_{i}\Sigma _{ij}^{L}\left( m_{i}^{2}\right) +m_{j}\Sigma
_{ij}^{L\dagger }\left( m_{j}^{2}\right) \right.  \nonumber \\
&&+m_{j}\Sigma _{ij}^{R}\left( m_{i}^{2}\right) +m_{i}\Sigma _{ij}^{R\dagger
}\left( m_{j}^{2}\right)  \nonumber \\
&&+m_{j}m_{i}\Sigma _{ij}^{\gamma R}\left( m_{i}^{2}\right)
+m_{i}m_{j}\Sigma _{ij}^{\gamma R\dagger }\left( m_{j}^{2}\right)  \nonumber
\\
&&+\left. m_{i}^{2}\Sigma _{ij}^{\gamma L}\left( m_{i}^{2}\right)
+m_{j}^{2}\Sigma _{ij}^{\gamma L\dagger }\left( m_{j}^{2}\right) \right\}
\quad \left( i\neq j\right) ,  \label{set2aherLij}
\end{eqnarray}
\begin{eqnarray}
\delta Z_{ii}^{R\dagger }+\delta Z_{ii}^{R} &=&m_{i}^{2}\left( \Sigma
_{ii}^{\gamma R\prime }\left( m_{i}^{2}\right) +\Sigma _{ii}^{\gamma R\prime
\dagger }\left( m_{i}^{2}\right) +\Sigma _{ii}^{\gamma L\prime }\left(
m_{i}^{2}\right) +\Sigma _{ii}^{\gamma L\prime \dagger }\left(
m_{i}^{2}\right) \right)  \nonumber \\
&&+m_{i}\left( \Sigma _{ii}^{R\prime }\left( m_{i}^{2}\right) +\Sigma
_{ii}^{R\prime \dagger }\left( m_{i}^{2}\right) +\Sigma _{ii}^{L\prime
}\left( m_{i}^{2}\right) +\Sigma _{ii}^{L\prime \dagger }\left(
m_{i}^{2}\right) \right)  \nonumber \\
&&+\Sigma _{ii}^{\gamma R}\left( m_{i}^{2}\right) +\Sigma _{ii}^{\gamma
R\dagger }\left( m_{i}^{2}\right) ,  \label{set2hermii}
\end{eqnarray}
and, in addition, those obtained after the replacement $R\leftrightarrow L$
The mass counterterm is identical to the one obtained in the incoming
fermion scheme.

Here again we obtain the relation (\ref{cons}). Note that diagonal
counterterms coincide in both schemes while this is not the case for
off-diagonal ones and of course when there is no mixing the usual
renormalization constants are reproduced.

So far we have presented the above schemes without specifying weak indices $%
u $ or $d$. In the next subsections we will see that the above schemes can
be imposed alternatively on up or down type fermions but not \emph{on both
at the same time}. The reason is that gauge symmetry impose certain
relations between renormalization constants that are not fulfilled in the
former case.

\subsection{The role of Ward identities}

Let us obtain the Ward identities that relate renormalization constants in
the physical base. The non-physical base belongs to an irreducible
representation of $SU_{L}\left( 2\right) $ (weak doublet) and we want the
renormalization group to respect this representation, that is 
\begin{eqnarray}
\mathrm{u}_{L}^{0} &=&Z^{L\frac{1}{2}}\mathrm{u}_{L},  \nonumber \\
\mathrm{d}_{L}^{0} &=&Z^{L\frac{1}{2}}\mathrm{d}_{L},  \label{eqa41}
\end{eqnarray}
where the wave function renormalization $Z^{L\frac{1}{2}}$ is the same for
the two components of the weak doublet. The non-physical basis is related to
the physical one via a bi-unitary transformation given by 
\begin{eqnarray}
\mathrm{u}_{L}^{0} &=&V_{Lu}^{0}u_{L}^{0},\quad \mathrm{u}_{L}=V_{Lu}u_{L}, 
\nonumber \\
\mathrm{d}_{L}^{0} &=&V_{Ld}^{0}d_{L}^{0},\quad \mathrm{d}_{L}=V_{Ld}d_{L},
\label{eqa42}
\end{eqnarray}
so we obtain 
\begin{eqnarray}
u_{L}^{0} &=&V_{Lu}^{0\dagger }Z^{L\frac{1}{2}}V_{Lu}u_{L}\equiv Z^{uL\frac{1%
}{2}}u_{L},  \nonumber \\
d_{L}^{0} &=&V_{Ld}^{0\dagger }Z^{L\frac{1}{2}}V_{Ld}d_{L}\equiv Z^{dL\frac{1%
}{2}}d_{L},  \label{renormrel}
\end{eqnarray}
where we have defined the wave function renormalization for the up and down
flavors in the physical basis as $Z^{uL\frac{1}{2}}=V_{Lu}^{0\dagger }Z^{L%
\frac{1}{2}}V_{Lu}$ and $Z^{dL\frac{1}{2}}=V_{Ld}^{0\dagger }Z^{L\frac{1}{2}%
}V_{Ld}$ respectively. From Eqs.(\ref{renormrel}) we immediately obtain \cite
{renormckm} 
\begin{equation}
K^{0}=V_{Lu}^{0\dagger }V_{Ld}^{0}=Z^{uL\frac{1}{2}}V_{Lu}^{\dagger
}V_{Ld}Z^{dL-\frac{1}{2}}=Z^{uL\frac{1}{2}}KZ^{dL-\frac{1}{2}},
\label{ward1}
\end{equation}
and 
\begin{eqnarray}
Z^{uL\dagger \frac{1}{2}}Z^{uL\frac{1}{2}} &=&V_{Lu}^{\dagger }Z^{L\dagger 
\frac{1}{2}}Z^{L\frac{1}{2}}V_{Lu}  \nonumber \\
&=&V_{Lu}^{\dagger }V_{Ld}Z^{dL\dagger \frac{1}{2}}Z^{dL\frac{1}{2}%
}V_{Ld}^{\dagger }V_{Lu}  \nonumber \\
&=&KZ^{dL\dagger \frac{1}{2}}Z^{dL\frac{1}{2}}K^{\dagger },  \label{ward2}
\end{eqnarray}
If we define the $CKM$ renormalization constant as $K^{0}=K+\delta K$ we can
rewrite Eqs. (\ref{ward1}) and (\ref{ward2}) in perturbation theory as 
\begin{eqnarray}
\delta K &=&\frac{1}{2}\left( \delta Z^{uL}K-K\delta Z^{dL}\right) ,
\label{ward1inf} \\
\delta Z^{uL\dagger }+\delta Z^{uL} &=&K\left( \delta Z^{dL\dagger }+\delta
Z^{dL}\right) K^{\dagger }.  \label{ward2inf}
\end{eqnarray}
Eqs. (\ref{ward1inf}) and (\ref{ward2inf}) relating renormalization
constants in the physical base are consequence of $SU_{L}\left( 2\right) $
gauge invariance and must be fulfilled by any renormalization scheme.

Now we can see that a simple solution to obtain all renormalization
constants respecting Ward identities is to impose one of the presented
on-shell schemes for the down (up) fermions and then use Eq.(\ref{ward2inf})
to obtain the left Hermitian part of the wave function for the up (down)
fermions. For the anti-Hermitian and right Hermitian parts of the up (down)
fermions we can use the same expressions used for the down (up), but with
the $u\leftrightarrow d$ replacement. This procedure leads to a finite set
of Green functions and it is obviously compliant with the Ward identities.
However, this procedure alone does not lead to and up and down propagator
with the desired properties listed in one of the two on-shell schemes. Thus
for external legs the above renormalization prescription must be
supplemented with an additional finite renormalization, ensuring the
compliance with the incoming or outgoing schemes (depending whether the
particle is in the in or out state). We will illustrate this point in the
next section where we calculate the contribution to the renormalization of
the $CKM$ matrix given by Eq.(\ref{ward1inf}) and the wave function
renormalization which in the effective Lagrangian comes in both cases solely
from $\mathcal{L}_{L}^{4}$.

\subsection{Contribution of $\mathcal{L}_{L}^{4}$ to wave-function
renormalization}

\label{dk}The operator $\mathcal{L}_{L}^{4}$ is the only one contributing to
self-energies and, hence, to the wave-function renormalization constants. It
also gives a contribution (among others) to the neutral current vertices
which (see Eq.(\ref{l7fv})), when compared to the tree level Standard Model
contribution, is proportional to 
\begin{equation}
\left[ \left( N^{4}+N^{4\dagger }\right) \tau ^{u}-\left( K^{\dagger }\left(
N^{4}+N^{4\dagger }\right) K\right) \tau ^{d}\right] L.  \label{eqa43}
\end{equation}
The contribution from $\mathcal{L}_{L}^{4}$ to the bare self energies is 
\begin{eqnarray}
\Sigma ^{R\left( u,d\right) } &=&\Sigma ^{L\left( u,d\right) }=0,  \nonumber
\\
\Sigma ^{\gamma Ru} &=&\Sigma ^{\gamma Rd}=0,  \nonumber \\
\Sigma ^{\gamma Ld} &=&K^{\dagger }\left( N^{4}+N^{4\dagger }\right) K, 
\nonumber \\
\Sigma ^{\gamma Lu} &=&-\left( N^{4}+N^{4\dagger }\right) ,  \label{eqa44}
\end{eqnarray}
hence using either the incoming or outcoming on-shell renormalization
conditions we obtain (both give identical results in the present case, but
note that this is not true in general) 
\begin{eqnarray}
\frac{1}{2}\left( \delta Z_{ij}^{dL}+\delta Z_{ij}^{dL\dagger }\right)
&=&\left( K^{\dagger }\left( N^{4}+N^{4\dagger }\right) K\right) _{ij},
\label{wfrd1} \\
\frac{1}{2}\left( \delta Z_{ij}^{dR}+\delta Z_{ij}^{dR\dagger }\right) &=&0,
\label{wfrd2} \\
\frac{1}{2}\left( \delta Z_{ij}^{dL}-\delta Z_{ij}^{dL\dagger }\right) &=&%
\frac{m_{j}^{d2}+m_{i}^{d2}}{m_{j}^{d2}-m_{i}^{d2}}\left( K^{\dagger }\left(
N^{4}+N^{4\dagger }\right) K\right) _{ij}\qquad \left( i\neq j\right) ,
\label{wfrd3} \\
\frac{1}{2}\left( \delta Z_{ij}^{dR}-\delta Z_{ij}^{dR\dagger }\right) &=&%
\frac{2m_{i}^{d}m_{j}^{d}}{m_{j}^{d2}-m_{i}^{d2}}\left( K^{\dagger }\left(
N^{4}+N^{4\dagger }\right) K\right) _{ij}\qquad \left( i\neq j\right) ,
\label{wfrd4} \\
\frac{1}{2}\left( \delta Z_{ii}^{dL}-\delta Z_{ii}^{dL\dagger }\right) &=&%
\frac{1}{2}\left( \delta Z_{ii}^{dR}-\delta Z_{ii}^{dR\dagger }\right) =0.
\label{wfrd5}
\end{eqnarray}
Had we have used the same conditions for the up fermions we would have
obtained 
\begin{eqnarray}
\frac{1}{2}\left( \delta Z_{ij}^{uL}+\delta Z_{ij}^{uL\dagger }\right)
&=&-\left( N^{4}+N^{4\dagger }\right) _{ij},  \label{wfru1} \\
\frac{1}{2}\left( \delta Z_{ij}^{uR}+\delta Z_{ij}^{uR\dagger }\right) &=&0,
\label{wfru2} \\
\frac{1}{2}\left( \delta Z_{ij}^{uL}-\delta Z_{ij}^{uL\dagger }\right) &=&-%
\frac{m_{j}^{u2}+m_{i}^{u2}}{m_{j}^{u2}-m_{i}^{u2}}\left( N^{4}+N^{4\dagger
}\right) _{ij}\qquad \left( i\neq j\right) ,  \label{wfru3} \\
\frac{1}{2}\left( \delta Z_{ij}^{uR}-\delta Z_{ij}^{uR\dagger }\right) &=&-%
\frac{2m_{i}^{u}m_{j}^{u}}{m_{j}^{u2}-m_{i}^{u2}}\left( N^{4}+N^{4\dagger
}\right) _{ij}\qquad \left( i\neq j\right) ,  \label{wfru4} \\
\frac{1}{2}\left( \delta Z_{ii}^{uL}-\delta Z_{ii}^{uL\dagger }\right) &=&%
\frac{1}{2}\left( \delta Z_{ii}^{uR}-\delta Z_{ii}^{uR\dagger }\right) =0,
\label{wfru5}
\end{eqnarray}
note that Eqs.(\ref{wfrd1}) and (\ref{wfru1}) are indeed incompatible with
the Ward identity (\ref{ward2inf}) as expected. A solution to this
incompatibility is simply to take one of the two sets as valid for one of
the fermions (or even none of them; for example we can use the minimal
scheme), use the Ward identity to determine the left Hermitian part of the
renormalization of the other fermion, while keeping the anti-Hermitian and
right Hermitian parts from the original prescription. The renormalization of
the $CKM$ matrix is then fixed by Eq.(\ref{ward1inf}). Then we proceed to
renormalize the external fermions with additional finite renormalization
constants $\hat{Z}^{uL\frac{1}{2}}$ and $\hat{Z}^{dL\frac{1}{2}}$ with $\hat{%
Z}^{uL\frac{1}{2}}Z^{uL\frac{1}{2}}$ and $\hat{Z}^{dL\frac{1}{2}}Z^{dL\frac{1%
}{2}}$ satisfying the incoming or outgoing schemes, as appropriate. For
instance a consistent scheme in the present case would be to retain Eqs. (%
\ref{wfrd1})-(\ref{wfrd5}), and then Eqs. (\ref{wfru2})-(\ref{wfru5}). Then
replace Eq.(\ref{wfru1}) by (\ref{ward2inf}), which implies 
\begin{equation}
\frac{1}{2}\left( \delta Z_{ij}^{uL}+\delta Z_{ij}^{uL\dagger }\right)
=\left( N^{4}+N^{4\dagger }\right) _{ij}.
\end{equation}
Note the sign difference with respect to Eq. (\ref{wfru1}).

The above one is a Ward identity-compliant set of wave function
renormalization constants. From them, it is immediate to read the way the $%
CKM$ matrix renormalizes. As for the additional (finite, if radiative
corrections were included) renormalization, in the present case this amounts
to 
\begin{eqnarray}
\delta \hat{Z}_{ij}^{dL} &=&0,  \nonumber \\
\frac{1}{2}\left( \delta \hat{Z}_{ij}^{uL}+\delta \hat{Z}_{ij}^{uL\dagger
}\right) &=&-2\left( N^{4}+N^{4\dagger }\right) _{ij},  \nonumber \\
\frac{1}{2}\left( \delta \hat{Z}_{ij}^{uL}-\delta \hat{Z}_{ij}^{uL\dagger
}\right) &=&0,  \label{eqa46}
\end{eqnarray}
but the whole procedure is (for the external legs) equivalent to use
directly Eqs. (\ref{wfrd1})-(\ref{wfru5}) in the first place.

The bare kinetic term in the physical base in the Standard Model is given by 
\begin{eqnarray}
\mathcal{L}_{kin} &=&i\mathrm{\bar{f}}\gamma ^{\mu }\left\{ \partial _{\mu
}+i\frac{e}{s_{W}}\left( K\frac{\tau ^{-}}{2}W_{\mu }^{+}+K^{\dagger }\frac{%
\tau ^{+}}{2}W_{\mu }^{-}\right) L+ieQA_{\mu }\right.  \nonumber \\
&&+\left. i\frac{e}{c_{W}s_{W}}\left[ \left( \frac{\tau ^{3}}{2}%
-Qs_{W}^{2}\right) L-Qs_{W}^{2}R\right] Z_{\mu }+ig_{s}\frac{\mathbf{\lambda 
}}{2}\cdot G_{\mu }\right\} \mathrm{f.}  \label{eqa47}
\end{eqnarray}
To calculate the tree level contribution of $\mathcal{L}_{L}^{4}$ via this
term after renormalization we write it as 
\begin{eqnarray}
\mathcal{L}_{kin} &\rightarrow &i\mathrm{\bar{f}}\gamma ^{\mu }\left[ \left( 
\hat{Z}^{uL\dagger \frac{1}{2}}Z^{uL\dagger \frac{1}{2}}L+Z^{uR\dagger \frac{%
1}{2}}R\right) \tau ^{u}+\left( \hat{Z}^{dL\dagger \frac{1}{2}}Z^{dL\dagger 
\frac{1}{2}}L+Z^{dR\dagger \frac{1}{2}}R\right) \tau ^{d}\right]  \nonumber
\\
&&\times \left\{ \partial _{\mu }+i\frac{e}{s_{W}}\left( Z^{uL\frac{1}{2}%
}KZ^{dL\frac{-1}{2}}\frac{\tau ^{-}}{2}W_{\mu }^{+}+Z^{dL\dagger \frac{-1}{2}%
}K^{\dagger }Z^{uL\dagger \frac{1}{2}}\frac{\tau ^{+}}{2}W_{\mu }^{-}\right)
L+ieQA_{\mu }\right.  \nonumber \\
&&+\left. i\frac{e}{c_{W}s_{W}}\left[ \left( \frac{\tau ^{3}}{2}%
-Qs_{W}^{2}\right) L-Qs_{W}^{2}R\right] Z_{\mu }+ig_{s}\frac{\mathbf{\lambda 
}}{2}\cdot G_{\mu }\right\}  \nonumber \\
&&\times \left[ \left( Z^{uL\frac{1}{2}}\hat{Z}^{uL\frac{1}{2}}L+Z^{uR\frac{1%
}{2}}R\right) \tau ^{u}+\left( Z^{dL\frac{1}{2}}\hat{Z}^{dL\frac{1}{2}%
}L+Z^{dR\frac{1}{2}}R\right) \tau ^{d}\right] \mathrm{f,}  \label{eqa48}
\end{eqnarray}
where we have introduced the additional finite renormalization constants $%
\hat{Z}^{uL\frac{1}{2}}$ and $\hat{Z}^{dL\frac{1}{2}}$ necessary to avoid
mixing and maintain residue 1 in the propagators. We have also renormalized $%
K$ according to the Ward identity (\ref{ward1}). With the renormalization
constants taken into account we observe that the total contribution of $%
\mathcal{L}_{L}^{4}$ to the neutral current vertices vanishes. This is a
very non-trivial check of the whole procedure . Of course nothing prevents
the appearance of $N^{4}$ at higher orders when one, for instance, performs
loops with the effective operators. But this a purely academic question at
this point.

Finally let us see what happens to the charged current vertices. The total
contribution of $\mathcal{L}_{kin}$ and $\mathcal{L}_{L}^{4}$ including
renormalization constants to the charged vertex is 
\begin{eqnarray}
&&\left( I+\frac{\delta \hat{Z}^{uL}+\delta \hat{Z}^{uL\dagger }}{4}-\frac{%
\delta \hat{Z}^{uL}-\delta \hat{Z}^{uL\dagger }}{4}\right) \left( I+\frac{%
\delta Z^{uL}+\delta Z^{uL\dagger }}{2}+\left( N^{4}-N^{4\dagger }\right)
\right)  \nonumber \\
&&\times K\left( I+\frac{\delta \hat{Z}^{dL}+\delta \hat{Z}^{dL\dagger }}{4}+%
\frac{\delta \hat{Z}^{dL}-\delta \hat{Z}^{dL\dagger }}{4}\right)  \nonumber
\\
&=&\left( I+\frac{\delta \hat{Z}^{uL}+\delta \hat{Z}^{uL\dagger }}{4}-\frac{%
\delta \hat{Z}^{uL}-\delta \hat{Z}^{uL\dagger }}{4}+\frac{\delta
Z^{uL}+\delta Z^{uL\dagger }}{2}+\left( N^{4}-N^{4\dagger }\right) \right) K
\nonumber \\
&&+K\left( \frac{\delta \hat{Z}^{dL}+\delta \hat{Z}^{dL\dagger }}{4}+\frac{%
\delta \hat{Z}^{dL}-\delta \hat{Z}^{dL\dagger }}{4}\right)  \nonumber \\
&=&\left( I+\left( N^{4}-N^{4\dagger }\right) -\frac{\delta \hat{Z}%
^{uL}-\delta \hat{Z}^{uL\dagger }}{4}\right) K+K\left( \frac{\delta \hat{Z}%
^{dL}-\delta \hat{Z}^{dL\dagger }}{4}\right)  \nonumber \\
&=&K+\left( N^{4}-N^{4\dagger }\right) K,  \label{eqa49}
\end{eqnarray}
where we have used the Ward identity (\ref{ward2inf}) along with Eqs. (\ref
{wfrd1}-\ref{wfrd5}), Eqs.(\ref{wfru2}-\ref{wfru5}) and Eq. (\ref{eqa46}).
We observe that the total contribution of $\mathcal{L}_{kin}+\mathcal{L}%
_{L}^{4}$ is in fact equal to the contribution of $\mathcal{L}_{L}^{4}$
alone. The contributions coming from the wave function and $CKM$
renormalizations cancel out at tree level. Another point to note is that
this particular contribution preserves the perturbative unitarity of $K$, in
accordance with the equations-of-motion argument. This completes the
theoretical analysis of the $CKM$ and wave-function renormalization.

\section{Some examples}

\label{examples}Let us now try to get a feeling about the order of magnitude
of the coefficients of the effective Lagrangian. We shall consider two
examples: the effective theory induced by the integration of a heavy doublet
and the Standard Model itself in the limit of a heavy Higgs.

In the heavy doublet case we shall make use of some recent work by Del
Aguila and coworkers\cite{paco}. These authors have recently analyzed the
effect of integrating out heavy matter fields in different representations.
For illustration purposes we shall only consider the doublet case here. As
emphasized in \cite{paco} while additional chiral doublets are surely
excluded by the LEP data, vector multiplets are not.

Let us assume that the Standard Model is extended with a doublet of heavy
fermions $Q$ of mass $M$, with vector coupling to the gauge field. For the
time being we shall assume a light Higgs. In addition there will be an
extended Higgs-Yukawa term of the form 
\begin{equation}
\lambda _{j}^{\left( u\right) }\bar{Q}\tilde{\phi}R\mathrm{u}_{j}+\lambda
_{j}^{\left( d\right) }\bar{Q}\phi R\mathrm{d}_{j},  \label{eqa50}
\end{equation}
where 
\begin{equation}
\phi =\frac{1}{\sqrt{2}}\left( 
\begin{array}{c}
\varphi _{1}+i\varphi _{2} \\ 
v+h+i\varphi _{3}
\end{array}
\right) ,\qquad \tilde{\phi}=i\mathbf{\tau }^{2}\phi ^{\ast },\qquad \mathrm{%
f}=\left( 
\begin{array}{c}
\mathrm{u} \\ 
\mathrm{d}
\end{array}
\right) .  \label{eqb51}
\end{equation}

The heavy doublet can be exactly integrated. This procedure is described in
detail in \cite{paco}. After this operation we generate the following
effective couplings (all of them corresponding to operators of dimension
six) 
\begin{eqnarray}
&&i\phi ^{\dagger }D_{\mu }\phi \mathrm{\bar{f}}\alpha _{\phi q}^{\left(
1\right) }\gamma ^{\mu }L\mathrm{f}+h.c.,  \nonumber \\
&&i\phi ^{\dagger }\mathbf{\tau }^{j}D_{\mu }\phi \mathrm{\bar{f}}\alpha
_{\phi q}^{\left( 3\right) }\gamma ^{\mu }\mathbf{\tau }^{j}L\mathrm{f}+h.c.,
\nonumber \\
&&i\phi ^{\dagger }D_{\mu }\phi \mathrm{\bar{f}}\alpha _{\phi u}\gamma ^{\mu
}\mathbf{\tau }^{u}R\mathrm{f}+h.c.,  \nonumber \\
&&i\phi ^{\dagger }D_{\mu }\phi \mathrm{\bar{f}}\alpha _{\phi d}\gamma ^{\mu
}\mathbf{\tau }^{d}R\mathrm{f}+h.c.,  \nonumber \\
&&\frac{1}{\sqrt{2}}\phi ^{t}\mathbf{\tau }^{2}D_{\mu }\phi \mathrm{\bar{f}}%
\alpha _{\phi \phi }\gamma ^{\mu }\mathbf{\tau }^{-}R\mathrm{f}+h.c., 
\nonumber \\
&&-\phi ^{\dagger }\phi \mathrm{\bar{f}}\tilde{\phi}\alpha _{u\phi }R\mathrm{%
u}+h.c.,  \nonumber \\
&&-\phi ^{\dagger }\phi \mathrm{\bar{f}}\phi \alpha _{d\phi }R\mathrm{d}%
+h.c.,  \label{eqa51}
\end{eqnarray}
where 
\begin{equation}
D_{\mu }\phi =\left( \partial _{\mu }+ig\frac{\mathbf{\tau }}{2}\cdot W_{\mu
}+i\frac{g^{\prime }}{2}B_{\mu }\right) \phi .  \label{eqa52}
\end{equation}
The coefficients appearing in (\ref{eqa51}) take the values 
\begin{eqnarray}
\alpha _{\phi q}^{\left( 1\right) } &=&0,  \nonumber \\
\alpha _{\phi q}^{\left( 3\right) } &=&0,  \nonumber \\
\left( \alpha _{\phi u}\right) _{ij} &=&-\frac{1}{2}\lambda _{i}^{\left(
u\right) \dagger }\lambda _{j}^{\left( u\right) }\frac{1}{M^{2}},  \nonumber
\\
\left( \alpha _{\phi d}\right) _{ij} &=&\frac{1}{2}\lambda _{i}^{\left(
d\right) \dagger }\lambda _{j}^{\left( d\right) }\frac{1}{M^{2}},  \nonumber
\\
\left( \alpha _{\phi \phi }\right) _{ij} &=&\frac{1}{2}\lambda _{i}^{\left(
u\right) \dagger }\lambda _{j}^{\left( d\right) }\frac{1}{M^{2}},  \nonumber
\\
\tilde{y}_{u} &\rightarrow &\tilde{y}_{u}\left( I-\alpha _{\phi
u}M^{2}\right) ,  \nonumber \\
\tilde{y}_{d} &\rightarrow &\tilde{y}_{d}\left( I+\alpha _{\phi
d}M^{2}\right) ,  \label{eqa53}
\end{eqnarray}

The above results are taken from \cite{paco} and have been derived in a
linear realization of the symmetry group, where the Higgs field, $h$, is
explicitly included, along with the Goldstone bosons. It is easy however to
recover the leading contribution to the coefficients of our effective
operators (\ref{effec}). The procedure would amount to integrating out the
Higgs field, of course. This would lead to two type of contributions:
tree-level and one loop. The latter are enhanced by logs of the Higgs mass,
but suppressed by the usual loop factor $1/16\pi ^{2}$. In addition there
are the multiplicative Yukawa couplings. It is not difficult to see though
that only the light fermion Yukawa couplings appear and hence the loop
contribution is small. To retain the tree-level contribution only we simply
replace $\phi $ by its vacuum expectation value.

Since $\alpha _{\phi q}^{(1)}$ and $\alpha _{\phi q}^{(3)}$ are zero there
is no net contribution to the left effective couplings. On the contrary, $%
\alpha _{\phi u},$ $\alpha _{\phi d},$ and $\alpha _{\phi \phi }$ contribute
to the effective operators containing right-handed fields 
\begin{eqnarray}
\frac{M_{R}^{2\dagger }+M_{R}^{2\dagger }}{2} &=&-\frac{v^{2}}{8}\left(
\alpha _{\phi d}+\alpha _{\phi d}^{\dagger }+\alpha _{\phi u}+\alpha _{\phi
u}^{\dagger }\right) ,  \nonumber \\
\frac{M_{R}^{2}-M_{R}^{2\dagger }}{2} &=&\frac{v^{2}}{8}\left( \alpha _{\phi
\phi }-\alpha _{\phi \phi }^{\dagger }\right) ,  \nonumber \\
\frac{M_{R}^{1}+M_{R}^{1\dagger }}{2} &=&\frac{v^{2}}{16}\left( \alpha
_{\phi d}+\alpha _{\phi d}^{\dagger }-\alpha _{\phi u}-\alpha _{\phi
u}^{\dagger }+\alpha _{\phi \phi }+\alpha _{\phi \phi }^{\dagger }\right) , 
\nonumber \\
\frac{M_{R}^{3}+M_{R}^{3\dagger }}{2} &=&\frac{v^{2}}{16}\left( \alpha
_{\phi d}+\alpha _{\phi d}^{\dagger }-\alpha _{\phi u}-\alpha _{\phi
u}^{\dagger }-\alpha _{\phi \phi }-\alpha _{\phi \phi }^{\dagger }\right) ,
\label{eqa54}
\end{eqnarray}
In the process of integrating out the heavy fermions new mass terms have
been generated, so the mass matrix (of the light fermions) needs a further
re-diagonalization. This is quite standard and can be done by using the
formulae given in section 3. After diagonalization we should just replace $%
M_{R}^{i}\rightarrow \tilde{M}_{R}^{i}$ and this is the final result in the
physical basis. As we can see, the contribution to the effective couplings,
and hence to the observables, is always suppressed by a power of $M^{-2}$,
the scale of the new physics, as announced in the introduction. The
contribution from many other models involving heavy fermions can be deduced
from \cite{paco} in a similar way and general patterns inferred.

The second example we would like to briefly discuss is the Standard Model
itself. Particularly, the Standard Model in the limit of a heavy Higgs. In
the case without mixing the effective coefficients were derived in \cite
{nomix}. The results in the general case where mixing is present are given
by 
\begin{eqnarray}
\left( \tilde{M}^{2}-\tilde{M}^{2\dagger }\right) _{ij} &=&-\frac{1}{16\pi
^{2}}\frac{m_{i}^{u}K_{ij}m_{j}^{d}-m_{i}^{d}K_{ij}^{\dagger }m_{j}^{u}}{%
4v^{2}}(\frac{1}{\hat{\epsilon}}-\log \frac{M_{H}^{2}}{\mu ^{2}}+\frac{5}{2}%
),  \nonumber \\
\left( \tilde{M}^{2}+\tilde{M}^{2\dagger }\right) _{ij} &=&\frac{1}{16\pi
^{2}}\frac{m_{i}^{d2}-m_{i}^{u2}}{4v^{2}}\left( \frac{1}{\hat{\epsilon}}%
-\log \frac{M_{H}^{2}}{\mu ^{2}}+\frac{5}{2}\right) \delta _{ij},  \nonumber
\\
\left( \tilde{M}^{1}+\tilde{M}^{1\dagger }\right) _{ij} &=&-\frac{1}{16\pi
^{2}}\frac{\left( m_{i}^{u2}+m_{i}^{d2}\right) \delta
_{ij}+m_{i}^{u}K_{ij}m_{j}^{d}+m_{i}^{d}K_{ij}^{\dagger }m_{j}^{u}}{8v^{2}}(%
\frac{1}{\hat{\epsilon}}-\log \frac{M_{H}^{2}}{\mu ^{2}}+\frac{5}{2}), 
\nonumber \\
\left( \tilde{M}^{3}+\tilde{M}^{3\dagger }\right) _{ij} &=&-\frac{1}{16\pi
^{2}}\frac{\left( m_{i}^{u2}+m_{i}^{d2}\right) \delta
_{ij}-m_{i}^{u}K_{ij}m_{j}^{d}-m_{i}^{d}K_{ij}^{\dagger }m_{j}^{u}}{8v^{2}}(%
\frac{1}{\hat{\epsilon}}-\log \frac{M_{H}^{2}}{\mu ^{2}}+\frac{5}{2}), 
\nonumber \\
\left( N^{4}+N^{4\dagger }\right) _{ij} &=&\frac{1}{16\pi ^{2}}\frac{%
m_{i}^{u2}\delta _{ij}-K_{ik}m_{k}^{d2}K_{kj}^{\dagger }}{4v^{2}}(\frac{1}{%
\hat{\epsilon}}-\log \frac{M_{H}^{2}}{\mu ^{2}}+\frac{1}{2}),  \nonumber \\
\left( N^{2\dagger }+N^{2}\right) _{ij} &=&\frac{1}{16\pi ^{2}}\frac{%
m_{i}^{u2}\delta _{ij}-K_{ik}m_{k}^{d2}K_{kj}^{\dagger }}{4v^{2}}\left( 
\frac{1}{\hat{\epsilon}}-\log \frac{M_{H}^{2}}{\mu ^{2}}+\frac{5}{2}\right) ,
\nonumber \\
\left( N^{1}+N^{1\dagger }\right) _{ij} &=&-\frac{1}{16\pi ^{2}}\frac{%
m_{i}^{u2}\delta _{ij}+K_{ik}m_{k}^{d2}K_{kj}^{\dagger }}{4v^{2}}(\frac{1}{%
\hat{\epsilon}}-\log \frac{M_{H}^{2}}{\mu ^{2}}+\frac{5}{2}),  \nonumber \\
\left( N^{3}+N^{3\dagger }\right) _{ij} &=&0,  \nonumber \\
\left( N^{2}-N^{2\dagger }\right) _{ij} &=&-\left( N^{4}-N^{4\dagger
}\right) _{ij},  \label{effmix}
\end{eqnarray}
where we have used dimensional regularization with $d=4-2\epsilon $ and $%
\left\{ \gamma ^{5},\gamma ^{\mu }\right\} =0$; we have also defined $\frac{1%
}{\hat{\epsilon}}=\frac{1}{\epsilon }-\gamma +\log 4\pi $. Form Eqs.(\ref
{effmix}), (\ref{Zcoupl}) and (\ref{vertices}) we immediately obtain the
contribution to the $Z$ and $W$ current vertices

\begin{eqnarray}
g_{L}^{u} &=&\frac{1}{16\pi ^{2}}\frac{m_{i}^{u2}\delta _{ij}}{2v^{2}}\left( 
\frac{1}{\hat{\epsilon}}-\log \frac{M_{H}^{2}}{\mu ^{2}}+\frac{5}{2}\right) ,
\nonumber \\
g_{L}^{d} &=&-\frac{1}{16\pi ^{2}}\frac{m_{i}^{d2}\delta _{ij}}{2v^{2}}(%
\frac{1}{\hat{\epsilon}}-\log \frac{M_{H}^{2}}{\mu ^{2}}+\frac{5}{2}), 
\nonumber \\
g_{R}^{u} &=&-\frac{1}{16\pi ^{2}}\frac{m_{i}^{u2}\delta _{ij}}{2v^{2}}(%
\frac{1}{\hat{\epsilon}}-\log \frac{M_{H}^{2}}{\mu ^{2}}+\frac{5}{2}), 
\nonumber \\
g_{R}^{d} &=&\frac{1}{16\pi ^{2}}\frac{m_{i}^{d2}\delta _{ij}}{2v^{2}}\left( 
\frac{1}{\hat{\epsilon}}-\log \frac{M_{H}^{2}}{\mu ^{2}}+\frac{5}{2}\right) ,
\nonumber \\
h_{L} &=&\frac{1}{16\pi ^{2}}\frac{m_{i}^{u2}K_{ij}+K_{ij}m_{j}^{d2}}{4v^{2}}%
(\frac{1}{\hat{\epsilon}}-\log \frac{M_{H}^{2}}{\mu ^{2}}+\frac{5}{2}), 
\nonumber \\
h_{R} &=&-\frac{1}{16\pi ^{2}}\frac{m_{i}^{u}K_{ij}m_{j}^{d}}{2v^{2}}(\frac{1%
}{\hat{\epsilon}}-\log \frac{M_{H}^{2}}{\mu ^{2}}+\frac{5}{2}).
\label{higgscoupl}
\end{eqnarray}
These coefficients summarize the non-decoupling effects of a heavy Higgs in
the Standard Model. Note that a heavy Higgs gives rise to radiative
corrections that do not contribute to flavor changing neutral currents, but
generates contributions to the charged currents that alter the unitarity of
the left mixing matrix $\mathcal{U}$ and produces a right mixing matrix
which is non-unitary and of course is not present at tree level.

The divergence of these coefficients just reflect that the Higgs is a
necessary ingredient for the Standard Model to be renormalizable. These
divergences cancel the singularities generated by radiative corrections in
the light sector. At the end of the day, this amounts to cancelling all $%
\frac{1}{\epsilon }$ and replacing $\mu \rightarrow M_{W}$.

Although, strictly speaking, the above results hold in the minimal Standard
Model, experience from a similar calculation (without mixing) in the
two-Higgs doublet model\cite{espriu-ciafaloni} leads us to conjecture that
they also hold for a large class of extended scalar sectors, provided that
all other scalar particles in the spectrum are made sufficiently heavy.
Unless some additional $CP$ violation is included in the two-doublet
potential, there is only one phase: the one of the Standard Model.

Thus we have seen how different type of theories lead to a very different
pattern for the coefficients of the effective theory and, eventually, to the 
$CP$-violating observables. Theories with scalars are, generically,
non-decoupling, with large logs, which are nevertheless suppressed by the
usual loop factors. Theories with additional fermions are decoupling, but
provide contributions already at tree level. For heavy doublets only in the
right-handed sector, it turns out.

\section{Conclusions}

\label{conclusions}In this work we have performed a rather detailed analysis
of the issue of possible departures from the Standard Model in the charged
current sector, with an special interest in the issue of possible new
sources of $CP$ violation. The analysis we have performed is rather general.
We only assume that all ---so far--- undetected degrees of freedom are heavy
enough for an expansion in inverse powers of their mass to be justified.

We have retained in all cases the leading contribution to the observables
from the effective Lagrangian. To be fully consistent one should, at the
same time, include the one-loop corrections from the Standard Model without
Higgs (universal). We have not done so, so our results are sensitive to the
contribution from the new physics ---encoded in the coefficients of the
effective Lagrangian--- inasmuch as this dominates over the Standard Model
radiative corrections. Anyhow, it is usually possible to treat radiative
corrections with the help of effective couplings, thus falling back again in
an effective Lagrangian treatment.

There are two main theoretical results presented in this work. First of all,
we have performed a complete study of all the possible new operators, to
leading order, and the way to implement the passage to the physical basis
when these additional interactions are included. To our knowledge this is
the first time that this issue has been considered in the present framework
with such an exhaustive detail. Secondly, we have analyzed in detail the
issue of wave function and $CKM$ matrix element renormalization. Both need
to be included when the contribution from the effective operators to the
different observables is considered. This has been, to our knowledge, been
ignored in past treatments in the literature. As mentioned in the paper, the
issue is interesting by itself.

We have also computed the relevant coefficients in a number of theories.
Theories with extended matter sectors give, in principle, relatively large
contributions, since they contribute at the tree level. When only heavy
doublets are considered, the relevant left couplings are left untouched.
Observable effects should be sought after in the right-handed sector. The
contribution from the new physics is decoupling (i.e. vanishes when the
scale is sent to infinity). However the limits on additional vector
generations are weak, roughly one requires only their mass to be heavier
than the top one, so this may lead to large contributions. Of course, there
are mixing parameters $\lambda $, which can be bound from flavour changing
phenomenology. Measuring the right-handed couplings seems the most promising
way to test these possible effects. Stringent bounds exist in this respect
from $b\rightarrow s\gamma $, constraining the couplings at the few per
mille level \cite{cleo}. If one assumes some sort of naturality argument for
the scale of the coefficients in the effective Lagrangian, this precludes
observation unless at least the $1\%$ level of accuracy is reached. Theories
with extended scalar sectors are (unless fine tuning of the potential is
present such as in e.g. supersymmetric theories) non-decoupling and in order
to make its contribution larger than the universal radiative corrections one
requires a heavy Higgs (although their contribution, with respect to
universal radiative corrections is nevertheless enhanced by the top Yukawa
coupling).

In general, even if the physics responsible for the generation of the
additional effective operators is $CP$-conserving, phases which are present
in the Yukawa and kinetic couplings become observable. This should produce a
wealth of phases and new $CP$- violating effects. As we have seen,
contributions reaching the $1\%$ level are not easy to find, so it will be
extremely difficult to find any sizeable deviation with respect the Standard
Model in the ongoing experiments.

A systematic study of the phenomenology of these couplings is now under way,
as is clear that a lot of work remains to be done, such as identifying the
adequate observables for the wealth of phases that might appear.
Furthermore, we have obtained the effective Lagrangian at the $M_{W}$ scale
and we still have to scale down to the $b$, $c$ or kaon mass, which is a non
trivial task.

On a more practical level our results are relevant on three different
fronts. First of all we have, hopefully, clarified the issue of
wave-function and $CKM$ matrix elements renormalization. While the use we
have made of our proposal is limited (only one coefficient of the effective
Lagrangian contributes to the wave function and $CKM$ renormalizations), our
proposal meets all the necessary requirements. Secondly, we can incorporate
a good part of the radiative corrections in the Standard Model itself in the 
$d=4$ effective operators (we have seen that explicitly for the Higgs
contribution) so our results will be relevant the day that experiments
become accurate enough so that radiative corrections are required. Finally,
the effective Lagrangian approach consists not only in writing down the
Lagrangian itself, but it comes with a well defined set of counting rules.
This set of counting rules allows in the case of the $CKM$ matrix elements a
perturbative treatment of the unitarity constraint. If one assumes that the
contribution from new physics and radiative corrections are comparable, then
it is legitimate to use the unitarity relations in all one-loop
calculations. On the contrary the tree-level predictions should be modified
to account for the presence of the new-physics which introduces new phases.
This procedure can be extended to arbitrary order.

\section{Acknowledgments}

J.M. acknowledges a fellowship from Generalitat de Catalunya, grant
1998FI-00614. Financial support from grants AEN98-0431, 2000SGR 00026 and
EURODAPHNE is greatly appreciated. G.D'Ambrosio, C.P.Burgess, F.del Aguila,
M.J.Herrero and J.Matias are thanked for many discussions.

\end{document}